\documentclass[]{bytedance_seed}
\usepackage[toc,page,header]{appendix}
\usepackage{minitoc}
\usepackage{makecell}
\usepackage{amssymb}
\usepackage{float}
\usepackage[numbers]{natbib}

\title{Accelerated Machine Learning Force Field for Predicting Thermal Conductivity of Organic Liquids}
\author[1, 2]{Wei Feng}
\author[1]{Siyuan Liu}
\author[1]{Hongyi Wang}
\author[1]{Zhenliang Mu}
\author[1]{Zhichen Pu}
\author[1]{Xu Han}
\author[1]{Tianze Zheng}
\author[1]{Zhenze Yang}
\author[1]{Zhi Wang}
\author[1]{Weihao Gao}
\author[2]{Yidan Cao}
\author[1, \dagger]{Kuang Yu}
\author[1, \dagger]{Sheng Gong}
\author[1]{Wen Yan}

\affiliation[1]{ByteDance Seed}
\affiliation[2]{Tsinghua Shenzhen International Graduate School}

\contribution[\dagger]{Corresponding authors}
\footnotetext[1]{Wei Feng and Siyuan Liu contributed equally to this work.}
\abstract{
The thermal conductivity of organic liquids is a vital parameter influencing various industrial and environmental applications, including energy conversion, electronics cooling, and chemical processing. However, atomistic simulation of thermal conductivity of organic liquids has been hindered by the limited accuracy of classical force fields and the huge computational demand of \textit{ab initio} methods. In this work, we present a machine learning force field (MLFF)-based molecular dynamics simulation workflow to predict the thermal conductivity of 20 organic liquids. Here, we introduce the concept of differential attention into the MLFF architecture for enhanced learning ability, and we use density of the liquids to align the MLFF with experiments. As a result, this workflow achieves a mean absolute percentage error of 14\% for the thermal conductivity of various organic liquids, significantly lower than that of the current \textit{off-the-shelf} classical force field (78\%). Furthermore, the MLFF is rewritten using Triton language to maximize simulation speed, enabling rapid prediction of thermal conductivity.
}

\date{\today}
\correspondence{Kuang Yu at \email{kuangyu.2025@bytedance.com}, Sheng Gong at \email{shenggongmit@gmail.com}}

\begin{document}
\maketitle

\clearpage

\section{Introduction}\label{intro}

The thermal conductivity of organic liquids is a critical parameter in numerous industrial and environmental applications. These liquids play essential roles in areas such as energy conversion, electronics cooling, and chemical processing. Understanding and optimizing the thermal conductivity of organic liquids can lead to more efficient thermal management systems~\cite{mitra2022cite01}, enhancing performance and sustainability. For instance, in the electronics industry, higher thermal conductivities of cooling fluids directly contribute to better heat dissipation, thereby extending device lifespans and reducing energy consumption~\cite{sanker2022cite02, yoo2007cite03}. In the context of renewable energy, organic heat transfer fluids with optimized thermal properties can significantly improve the efficiency of solar thermal and geothermal installations~\cite{franca2018cite04, van2005cite05, mohapatra2005cite06}. Furthermore, in the development of liquid electrolytes, heat transfer is closely related to the safety and longevity of lithium batteries \cite{wang2008electrolyte01, takami2001electrolyte02, ye2024electrolyte03}.

Despite the importance of the thermal conductivities of liquids, accurately simulating it using atomistic simulation remains challenging. Traditionally, there are two main approaches to simulate the thermal conductivity of liquids in terms of calculating atomic interactions. The first is to use classical force fields-based molecular dynamics (MD) simulations, which are thought to have limited accuracy due to the simple functional form and limited expressive power of classical force fields. For example, in \citet{zhang2005water_classical_force_field}, the simulated thermal conductivity of water is overestimated by 33\% via reverse non-equilibrium molecular dynamics (rNEMD) with a classical force field. Although the accuracy of predicting the thermal conductivity of water has been gradually improved by force fields specifically designed for water~\cite{galli}, tailoring classical force fields for a specific property of a liquid typically requires the input of the experimentally measured property of that liquid, and this approach often deteriorate the prediction of other properties~\cite{watermodel}. As a result, to date classical force fields have not been proved to be effective and reliable for simulating thermal conductivity of general molecular liquids. 

The second approach is using \textit{ab initio} molecular dynamics (AIMD) simulation. Although \textit{ab initio} simulations have the potential to accurately describe atomic interactions for most molecules, the quality of simulations depends on the specific form of the \textit{ab initio} method. For example, Tisi \textit{et al} \cite{tisi2021dp_heat} used AIMD based on the Perdew-Burke-Ernzerhof (PBE) density functional~\cite{becke1988gga_1, becke1997gga_2} and the strongly constrained and appropriately normed (SCAN) functional~\cite{sun2015scan_meta_gga} to simulate the thermal conductivity of water. The error of PBE-based simulation (60\% overestimation) is larger than that of the SCAN-based simulations (20\% overestimation). In addition to the dependence of accuracy on functional forms, the system size effect~\cite{sellan2010size_effect, chantrenne2004finite_size, muller1997rnemd, cheng2020size_eq} of the simulation of thermal conductivity requires large simulation box, which makes AIMD an inefficient approach for simulating thermal conductivity of liquids due to its high scaling of computational cost with respect to the system size ($\geq \mathcal{O}(N^3)$). Despite the generalizability of \textit{ab initio} simulations, there is still no study reporting that AIMD can predict thermal conductivity of water more accurately than classical force fields~\cite{galli}. 

Recently, machine learning force fields (MLFFs)~\cite{unke2021machine} have been increasingly used to perform MD simulations~\cite{frenkel2023understanding}, owing to MLFFs' ability to calculate interatomic interactions significantly faster than \textit{ab initio} simulations, while fiting \textit{ab initio} data with higher accuracy compared with classical force fields. Despite the widespread application of MLFFs on simulating various properties of solid-state materials~\cite{m3gnet, chgnet, genome, dpa2, macemp}, there are far fewer literature reports on the application of MLFFs on simulating properties of molecular liquids~\cite{maceoff23}, especially transport properties. For thermal conductivity of liquids, to the best of the authors' knowledge, water is the only system of which has been simulated by MLFF~\cite{tisi2021dp_heat, zhang2023dp_thermal}. This scarcity might come from the fact that, the accuracy of most MLFFs is still limited by the underlying \textit{ab initio} training data. Recently, reports have suggested the potential of finetuning MLFFs for simulating the thermal conductivity of organic polymer and other systems~\cite{phyneo,tu_enhancing_2025, zhuang1, zhuang2}. In general, to address this limitation, two strategies have been proposed. The first is to use CCSD(T)~\cite{purvis1982ccsd_t, raghavachari1989ccsdt} data, the golden standard quantum chemistry simulation method, to train or finetune the MLFF~\cite{smith2019approaching}. However, this approach requires substantial computational resources to generate the training data, and has not been proved to accurately simulate transport properties of organic liquids. The second strategy is using experimental data to align MLFFs with the real world~\cite{dmff,han2025refining,gong2024bamboo,feng_screening_2025}. In our recent work~\cite{gong2024bamboo,feng_screening_2025}, we have shown that, aligning the density from the MLFF-based MD simulations and experiments can lead to improved prediction of viscosity and ionic conductivity from the MLFF. This transferability between density and transport properties is inspiring, as density is a property widely available in experiments and easy to calculate in MD while transport properties are generally harder to measure in experiments and to converge in MD.

In this work, we build a MLFF-based MD workflow, BAMBOO-TC (\textbf{B}yteDance \textbf{A}I \textbf{M}olecular Simulation \textbf{Boo}ster-\textbf{T}hermal \textbf{C}onductivity), that can simulate thermal conductivities of 20 commonly used organic liquids with the average deviation of 14\% by a single MLFF model, which is significantly lower than that (78\%) of the current \textit{off-the-shelf} classical force field (OPLS-AA~\cite{oplsaa}) for organic liquids. Compared with BAMBOO~\cite{gong2024bamboo}, in BAMBOO-TC we introduce the concept of ``differential attention''\cite{ye2024differential} into the graph equivariant transformer architecture (GET) to improve the expressive power of the MLFF. Moreover, we find that aligning the density between MLFF-based MD and experiments can reduce the error of thermal conductivity prediction significantly (from 25\% to 14\%), which shows the necessity of alignment between MLFF and the real-world for simulating transport properties of organic liquids~\cite{gong2024bamboo}. Finally, to address the high computational cost associated with MLFF inference during molecular dynamics simulations, the BAMBOO-TC model was re-implemented in the Triton framework to maximize simulation speed, enabling rapid prediction of thermal conductivity.

\section{Result}\label{rslt}
\subsection{Workflow of BAMBOO-TC}\label{rslt_bamboo}

\begin{figure}
    \centering
    \includegraphics[width=1\linewidth]{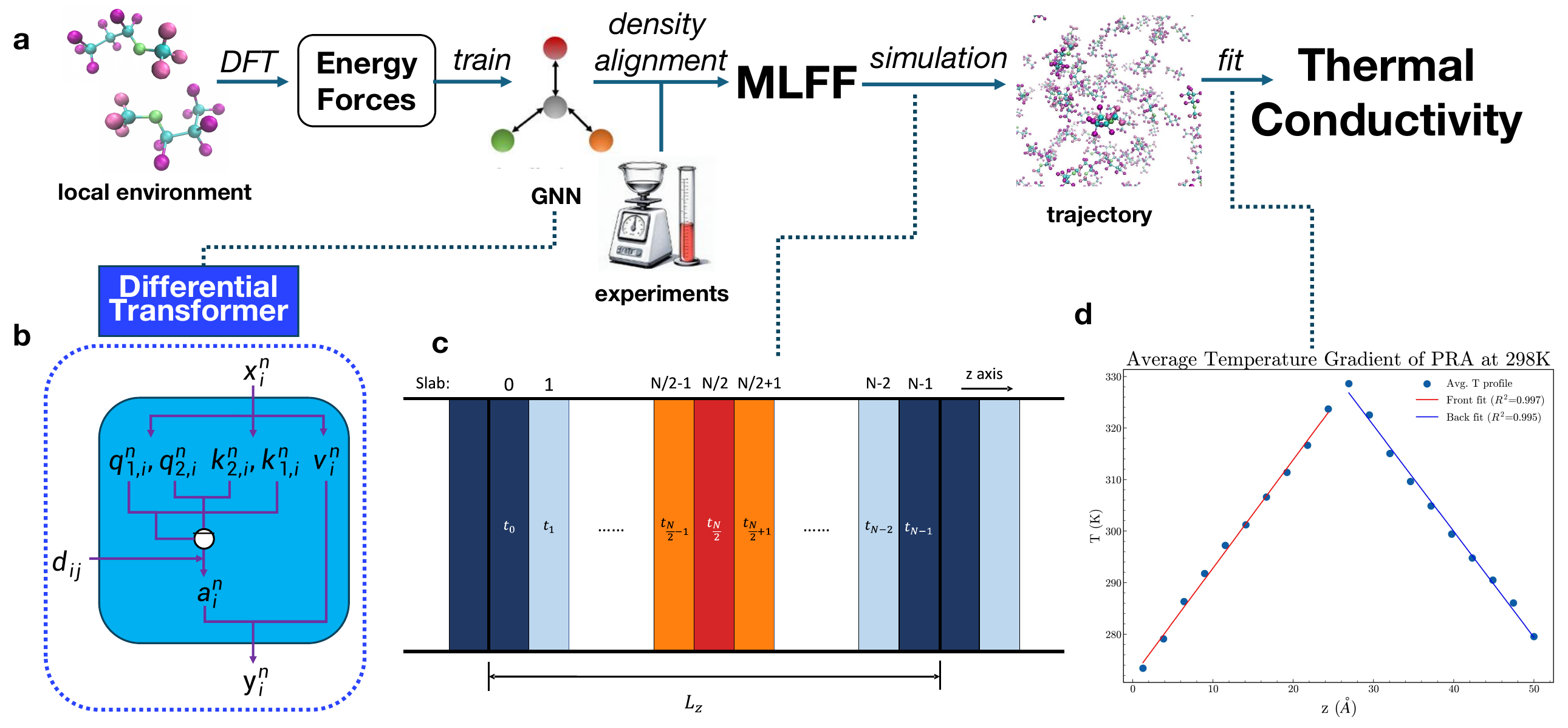}
    \caption{Overview of the MLFF-based workflow for prediction of thermal conductivity. \textbf{a}: Schematic of the sequential training and prediction processes of the workflow. \textbf{b}: Schematic of the differential transformer inside the GEDT layers of the MLFF. \textbf{c}: Illustration of the reverse non-equilibrium molecular dynamics (rNEMD) for the prediction of thermal conductivity. The box is divided into slabs along z-axis, where a temperature gradient is created. The center slab is created as the hottest, both ends are created as the coolest. \textbf{d}: The temperature profile of PRA at 298K, including the upper and the lower half of the box in the simulation of tert-butanol.}
    \label{fig:scheme}
\end{figure}

Here, we train the MLFF using B3LYP-level DFT data~\cite{b3lyp}, and align the MLFF with experimental data, as illustrated in Figure~\ref{fig:scheme}. First, we sample the local atomic environments in organic liquids as gas-phase clusters, and then use B3LYP functional~\cite{b3lyp} and def2-svpd basis set~\cite{svpd} to calculate their energies and atomic forces. Note that the B3LYP functional~\cite{b3lyp} is a hybrid-level density functional, which is considered more accurate than PBE~\cite{becke1988gga_1, becke1997gga_2} and SCAN~\cite{sun2015scan_meta_gga} used in Tisi \textit{et al.}~\cite{tisi2021dp_heat}. In this work, with the aim of showcasing the broad applicability of predicting the thermal conductivity of organic liquid systems, we include a variety of organic molecules in the DFT dataset, which can be categorized by functional groups. As shown in Figure~\ref{fig:benchmark}a, the selected molecules span across alcohols, esters and carbonates. These selected molecules cover a variety of application scenarios, including but not limited to electrolytes, organic solvents, coolant fluids, etc., demonstrating the model's generalizability across different industrial scenarios. After training on the DFT dataset, we align the MLFF model with experimentally measured density data, thereby reducing the systematic error between DFT and real world. More details regarding the training process are provided in the Supplementary Information.

In Figure~\ref{fig:scheme}b, we illustrate the Graph Equivariant Differential Transformer (GEDT) architecture employed in this work, which is built on the GET architecture in BAMBOO. Unlike conventional transformer~\cite{vaswani2017attention} where interaction between each pair of nodes is encoded by a single attention function, in the differential transformer~\cite{ye2024differential} correlations across the nodes are captured by computing the difference between two attention functions, preserving the correlations while eliminating the attention noise and improving the prediction performance. As illustrated in Table~\ref{tab:dft} and Figure Supplementary~\ref{fig:bamboo_dft}, the GEDT model has lower prediction error with respect to DFT compared to the GET model, demonstrating the effectiveness of the differential attention on improving the ability of MLFF to learn DFT data. Detailed equations for the GEDT architecture are provided in the Supplementary Information~\ref{subsec:si_gedt}.

\begin{table}
    \caption{Comparison between the Graph Equivariant Differential Transformer (GEDT) and Graph Equivariant Transformer (GET) for predicting DFT derived energy, forces, and virial tensors. Errors are calculated based on the test set.}
    \label{tab:dft}
    \centering
    \begin{tabular}{c|cccccc}
        \hline
        &   Energy R$^2$& \makecell{Energy RMSE \\ (kcal/mol)} & Forces R$^2$& \makecell{Forces RMSE \\ (kcal/mol*\AA)} & Virial R$^2$& \makecell{Virial RMSE \\ (kcal/mol)} \\
        \hline
        GEDT& \textbf{1.0000}& \textbf{2.5213}& \textbf{0.9981}& \textbf{1.2254}& \textbf{0.9990}& \textbf{5.9833}\\
        GET& 1.0000& 2.9041& 0.9978& 1.3277& 0.9989&6.2993\\
        \hline 
    \end{tabular}
\end{table}

After the initial training, density alignment is conducted to calibrate the difference between the DFT training data and the real world. We apply the physics-grounded and efficient alignment process in BAMBOO~\cite{gong2024bamboo} to ensure its effectiveness and transferability. Comparing with the previous work~\cite{xu2023nqe_water} that predict the thermal conductivity of water with MLFF, a manually assigned density is not required in this work. This relaxation of condition results in better consistency during MD simulations. Experimental density can be used to determine the pressure adjustments necessary to align MD simulations with the experimental density. Subsequently, we can relate these pressure adjustments to inter-molecular forces, and then use the adjusted forces to refine the parameters of the MLFF. In systems where hydrogen bonds and light atoms are prevalent, e.g. water, density and thermal conductivity are both influenced by nuclear quantum effect (NQE)~\cite{xu2024nqe_density, luo2020nqe_ice} during MD calculations. However, it is usually weaker in most organic systems and we do not further elaborate on this aspect in the present work. In addition, the path integral molecular dynamics (PIMD) that properly considers NQE in molecular dynamics simulations is forbiddingly expensive~\cite{xu2024nqe_density}. Since density is the fundamental property of liquids that determines the inter-molecular distance and subsequently influence properties related to inter-molecular interaction such as viscosity and diffusivity~\cite{gong2024bamboo}, we expect that aligning density between MLFF and experiment would also decrease the gap between thermal conductivities from MLFF-based MD and experiments. Here we do not use thermal conductivity itself as the target to perform alignment for several reasons, because 1) the relation between thermal conductivity and pressure (the microscopic quantity to finetune the MLFF) is not as clear as that between density and pressure; 2) thermal conductivities measured by MD simulations is intrinsically noisy with fluctuations much larger than that of equilibrated density as in Figure~\ref{fig:thermal} (fluctuation of density in Figure~\ref{fig:thermal}a is much smaller than that of thermal conductivity in Figure~\ref{fig:thermal}b); 3) compared with thermal conductivity where few experimental measurements are available, experimental density data of organic liquids is more readily available. More details of the density alignment process can be found in Supplementary Information~\ref{subsec:si_alignment}.

\subsection{Predicting Thermal Conductivity of Organic Liquid Systems}\label{rslt_thermal}

\begin{figure}
    \centering
    \includegraphics[width=0.9\linewidth]{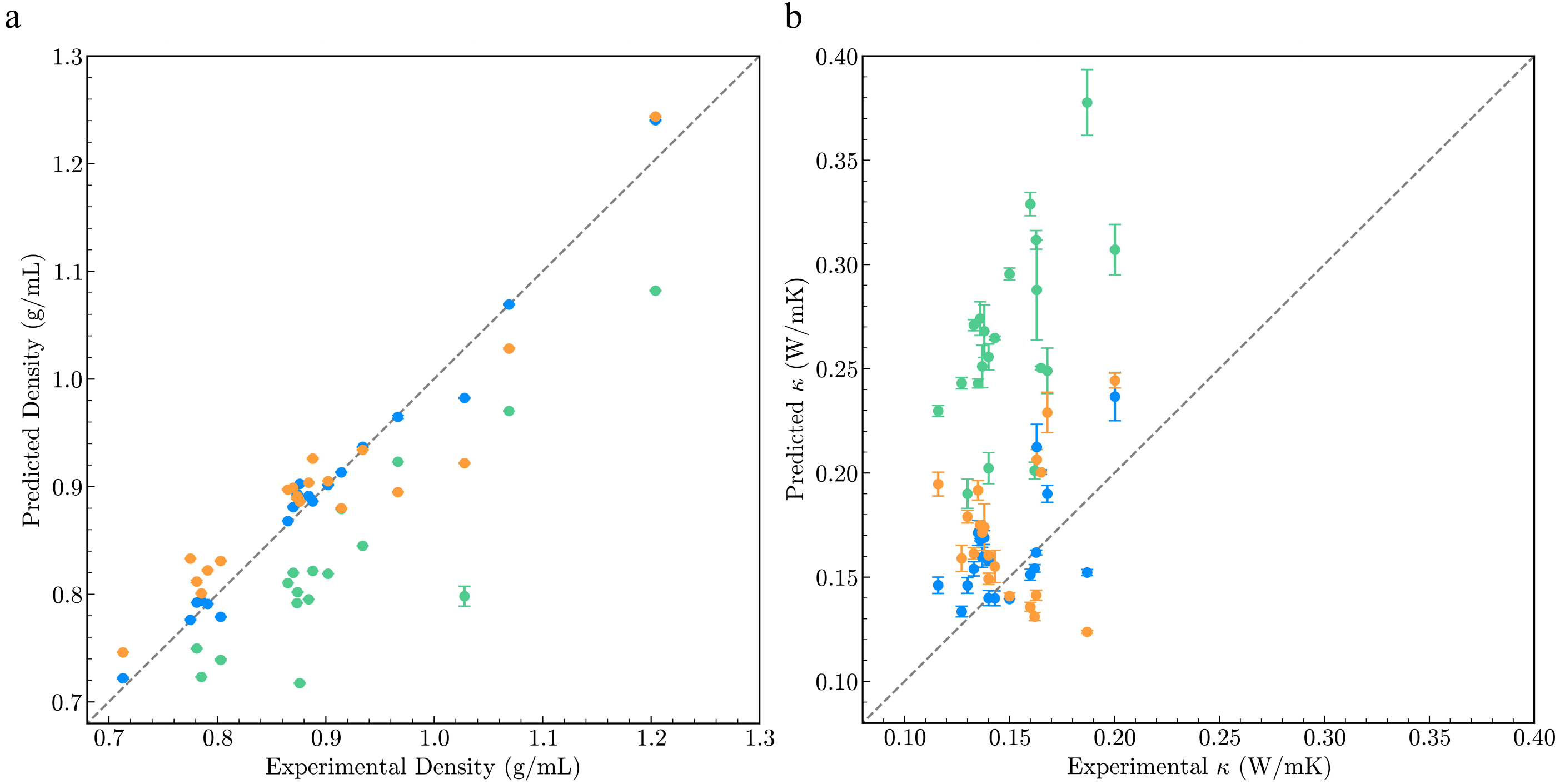}
    \caption{Prediction results of OPLS-AA and BAMBOO-TC before and after alignment. In both subfigures, predictions of OPLS-AA are shown in green circles, predictions of BAMBOO-TC before density alignment are shown in orange circles, predictions of BAMBOO-TC after density alignment are shown in blue circles. \textbf{a}: Correlation scatter comparing predicted densities to experimental densities. \textbf{b}: Correlation scatter comparing predicted thermal conductivities to experimental thermal conductivities. In all figures, the black line refers to $y=x$, and the standard deviation of the three independent replicates is used as the error bar.}
    \label{fig:thermal}
\end{figure}

We apply rNEMD~\cite{muller1997rnemd} to predict thermal conductivity as implemented in LAMMPS~\cite{thompson2022lammps}. We choose rNEMD because: 1) it avoids the calculation of heat fluxes and energy density which are non-trivial to define in MLFF~\cite{mlffheatflux,TORRESSANCHEZ2016224}; 2) its size convergence is better than other newly developed equilibrium methods~\cite{galli, cheng2020computing}. Here, we establish a temperature gradient along the z-direction, creating a controlled heat flow as shown in Figure~\ref{fig:scheme}c and d. We compute the thermodynamic temperature of the system by layers and average the temperature in these layers over specified time intervals to assess heat transport. From the size effect tests, we find that the thermal conductivity approximately converges when the simulation box contains about 12,000 atoms. Based on this, for each organic liquid system, we performed three fully independent simulations, including independent system construction, equilibration, and production rNEMD runs. The reported thermal conductivities are the mean values and the standard deviations over three independent simulations are used as error bars, as the final results shown in Figure~\ref{fig:thermal}.

\begin{table}
    \caption{Comparison of the prediction performance for density and thermal conductivity (Kappa) prediction from the OPLS-AA~\cite{oplsaa} classical empirical force field and BAMBOO-TC before and after density alignment. The prediction results are presented in mean absolute percentage error (MAPE) and R$^2$ score. }
    \label{tab:thermal}
    \centering
    \begin{tabular}{c|cccc}
        \hline
        Method&   Density RMSE& Density MAPE (\%)& Kappa RMSE& Kappa MAPE (\%) \\
        \hline
        \makecell[c]{OPLS-AA \\ }& 0.1025& 9.83\%& 0.1209& 78.01\%\\
        \makecell[c]{BAMBOO-TC \\ before alignment}& 0.0404& 3.65\%& 0.0407& 24.85\%\\
        \makecell[c]{BAMBOO-TC  \\ after alignment}& 0.0171& 1.25\%& 0.0252& 14.14\%\\
        \hline 
    \end{tabular}
\end{table}

In Table~\ref{tab:thermal} we show statistical results, including the RMSE and the mean absolute percentage error (MAPE), for predictions of density and thermal conductivity compared with experimental data. The density prediction from BAMBOO-TC demonstrates high accuracy, achieving a RMSE of 0.0171 and MAPE of 1.25\% . For thermal conductivity, predictions based on the rNEMD yield a RMSE of 0.0252 for BAMBOO-TC and 0.1209 for OPLS-AA, with corresponding MAPE of 14.14\% for BAMBOO-TC and 78.01\% for OPLS-AA. It is worth noting that the error in density and thermal conductivity predictions using the initially trained BAMBOO-TC model is greater than that of BAMBOO-TC after alignment, but outperforms OPLS-AA, with density MAPE of 3.65\% and thermal conductivity MAPE of 24.85\%, indicating the superiority of our framework and necessity of alignment with experimental data. Considering the mean absolute deviation for the thermal conductivity experimental measurement is 2.21\%~\cite{sykioti2013methanol_tc}, BAMBOO-TC not only significantly outperforms classical force fields in this regard but also sets a new benchmark for thermal conductivity predictions, effectively narrowing the significant gap between MD predictions and experiments. 

Additionally, the improvement of BAMBOO-TC's predictions after alignment also highlights the importance of introducing experimental data to fill the gap between B3LYP-level DFT calculation and experiment. Here we show that the B3LYP hybrid functional~\cite{b3lyp} along cannot lead to accurate thermal conductivity predictions, which is an extension to the previous observation~\cite{tisi2021dp_heat, galli} that GGA (PBE)~\cite{perdew1996pbe_gga} and meta GGA (SCAN)~\cite{sun2015scan_meta_gga} functionals cannot achieve the same degree of accuracy of thermal conductivity prediction for water as classical force fields. For future efforts, we recommend testing thermal conductivity of molecular liquids purely based on \textit{ab initio} methods, and we suggest to test the effect of training or finetuning the MLFF on higher level DFT data (such as those from double hybrid functionals~\cite{doublehybrid} and coupled-cluster methods~\cite{raghavachari1989ccsdt}).

Figure~\ref{fig:thermal} shows the predicted density and thermal conductivity of organic liquids using BAMBOO-TC and classical force fields against experimental data. Although all three models exhibit good overall correlations, with the two MLFFs showing higher predictive accuracy. However, the predictions of thermal conductivity in Figure~\ref{fig:thermal}b using OPLS-AA exhibit systematic overestimation, and BAMBOO-TC before alignment also shows overestimation for molecules with thermal conductivity less than 0.15 W$/$mK. The BAMBOO-TC model after density alignment corrects this overestimation and significantly improves the accuracy of thermal conductivity predictions, which demonstrates the advantage of MLFF over classical force fields for simulating thermal conductivity, and highlights the importance of alignment between the DFT-pretrained MLFF and the real world.

\begin{figure}
    \centering
    \includegraphics[width=0.99\linewidth]{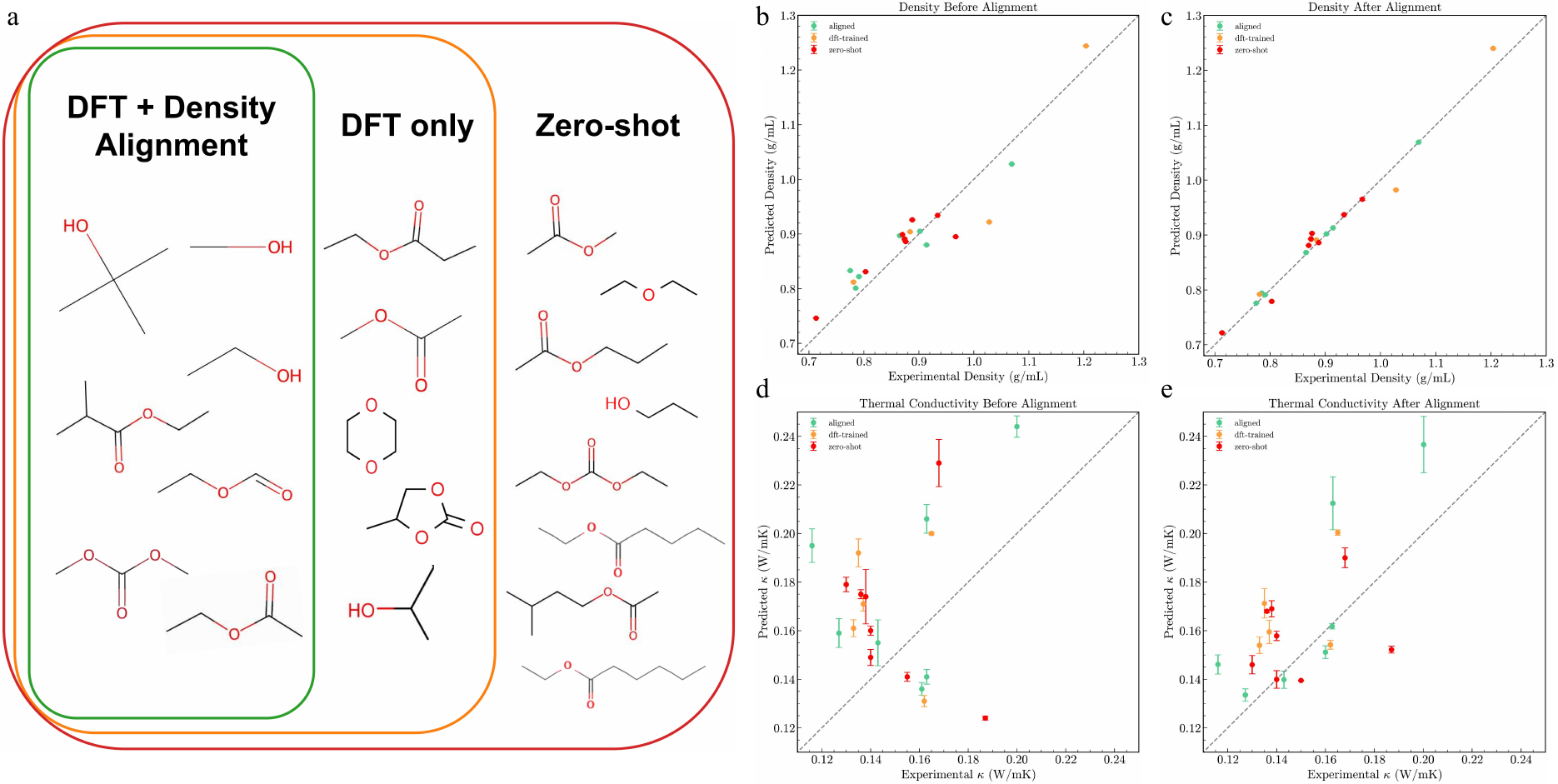}
    \caption{Assessment of transferability. \textbf{a}: Three categories of molecule: the BAMBOO-TC model presented in this work is initially trained with atomic clusters of \textit{aligned} and \textit{trained} molecules, and aligned with experimental data of \textit{aligned} molecules. \textit{Zero-shot} molecules do not participate in the training of BAMBOO-TC. \textbf{b,d}: Density (b) and thermal conductivity (d) predictions from BAMBOO-TC before density alignment. \textbf{c,e}: Density (c) and thermal conductivity (e) predictions from BAMBOO-TC after density alignment. In all figures, the black line refers to $y=x$, and the standard deviation of the three independent replicates is used as the error bar. }
    \label{fig:benchmark}
\end{figure}

In Figure~\ref{fig:benchmark}, we show the transferability of this workflow by categorizing molecules into three groups: aligned, DFT-trained, and zero-shot. This classification assesses the transferability of BAMBOO-TC outside DFT training data and densities used to align the MLFF. Figure~\ref{fig:benchmark}a illustrates the 20 molecules categorized accordingly. These 20 molecules were chosen based on the transferability of MLFFs to out-of-distribution molecules and the availability of experimental data. The selected molecules encompass a diverse array of organic liquids, including alcohols, esters and carbonates, which are widely utilized in applications such as liquid electrolytes and working fluids. This molecular set incorporates critical chemical motifs—hydrogen bonding, branching, and polarity—that govern thermal transport mechanisms in liquid-phase systems. Due to limitations in the availability of high-accuracy experimental data, the benchmark was refined to a curated set of 20 molecules, ensuring a representative balance across chemical functionality while preserving structural diversity. Of the 20 molecules in this study, 12 are used in the initial training of GEDT using DFT-calculated labels. Among these 12 DFT-trained molecules, the experimental density data of 7 molecules are additionally employed in aligning BAMBOO-TC.

Figures~\ref{fig:benchmark}b-e illustrate the comparison of predictive performance before and after the density alignment. Comparing Figure~\ref{fig:benchmark}b and c, we can see that after alignment, predictions of density for all the three types of molecules was improved, consistent with the observation in BAMBOO~\cite{gong2024bamboo}. For thermal conductivity, as in Figure~\ref{fig:benchmark}d and e, we can also see that, predictions for all the three categories correlate better to the experimental values after alignment, which suggests that the effect of density alignment is transferable to thermal conductivity of molecules that are unseen but similar to those included in the training set of DFT-based pretraining and density alignment. We acknowledge that, the transferability study here does not include molecules that are very different from the training set in terms of structural characteristics. In fact, we do not expect strong transferability of current MLFFs. As observed in Ref.~\cite{gong2024bamboo,maceoff23}, current MLFFs cannot guarantee stable MD simulations for arbitrary molecules even if the MLFFs are trained on DFT datasets with millions of quantum chemistry data samples. The studies in previous work~\cite{gong2024bamboo} and this section both suggest that transferability to out-of-distribution molecules is generally limited to molecules that contain only functional groups appearing in the DFT training dataset. Further efforts are still needed to improve the MLFFs to make them more transferable to out-of-sample systems.

\subsection{Accelerating Simulations with Triton}\label{subsec:rslt_triton}

While machine-learned force fields (MLFFs) offer superior accuracy over classical potentials, their high computational cost is a significant barrier to adoption. This performance bottleneck can slow simulations by orders of magnitude, limiting their application in demanding scenarios such as predicting transport properties. Although in our previous work BAMBOO~\cite{gong2024bamboo}, this has been mitigated by selecting the fast GET as the network architecture, the relatively slow simulation speed still remains a problem. Therefore, in BAMBOO-TC, we implement GEDT in Triton~\cite{tillet2019triton}, aiming to further accelerate model inference speed and eventually expand the domain of tractable simulations.

Our optimization strategy began by identifying the primary performance bottlenecks in the GEDT architecture: the radial basis function (RBF) layer and the attention layers. These layers are the most computationally intensive because they perform calculations for every atomic pair (i.e., at the edge level), whereas other layers operate on individual atoms (node level). We then re-implemented these layers as custom Triton kernels with a simple strategy: compute as many intermediate steps as possible on GPU registers and only access the memory when required. This is because the layers transform their inputs into their outputs with multiple steps, with each step reading its inputs from the GPU memory and writing their outputs back to the GPU memory even though the outputs of a step is the inputs of the next step, wasting most of the GPU memory bandwidth. Our strategy minimized GPU memory bandwidth usage and greatly reduces the running time of the kernels. Finally, since calculating atomic forces requires gradient information, a custom forward pass necessitates a corresponding custom backward pass. We therefore also implemented custom backward kernels in Triton for these layers to ensure their differentiability.

Cumulatively, compared to the scripted (\texttt{torch.jit.script}) version of the PyTorch model, the optimized model achieves an \textasciitilde 8x speedup for inference on an Nvidia H100 GPU. When we integrate the optimizations into a full simulation, this corresponds to an overall simulation speedup of \textasciitilde 3.7x. Performance comparisons on various GPU models can be found in~\autoref{tab:triton-speedup}. The discrepancy between the model inference and simulation speedups is attributed to computational overheads within the LAMMPS engine, the optimization of which is under active development but falls outside the scope of this work.

\begin{table}[ht]
\centering
\setlength{\tabcolsep}{10pt}
\renewcommand{\arraystretch}{1.25}
\caption{The speedups of using custom Triton kernels in molecular simulation. Unit for simulation speeds: ns/day. Note that the Triton launch meta-parameters are not tuned for the specific models, and further speedups can be expected if they are tuned to the optimal.}
\label{tab:triton-speedup}
\begin{tabular}{lccc}
\toprule
\textbf{GPU model} &
\textbf{Speed (w/o Triton)} &
\textbf{Speed (w/ Triton)} &
\textbf{Speedup} \\
\midrule
Nvidia V100   & 2.449 & 6.104 & 2.5x \\
Nvidia A10   & 1.637 & 5.641 & 3.5x \\
Nvidia H100   & 6.200 & 22.715 & 3.7x \\
\bottomrule
\end{tabular}
\end{table}

\section{Discussion}\label{disc}

In summary, this work demonstrates for the first time that a single MLFF model can be used to simulate thermal conductivity of 20 organic liquids with an average deviation of \textasciitilde 14\%. The proposed BAMBOO-TC workflow, built upon the graph equivariant differential transformer (GEDT) architecture, achieves significantly lower prediction errors of thermal conductivity compared with the OPLS-AA classical force field (\textasciitilde 78\% deviation). Our findings demonstrate that MLFFs can effectively enhance the accuracy of thermal conductivity predictions for organic liquids, thereby addressing a significant gap in the existing literature. We also highlight that, for the MLFF initially trained on hybrid-GGA level DFT (B3LYP) data, density alignment between the MLFF and the real-world can substantially reduce the prediction errors of density and thermal conductivity, which we hope can inspire further development on finetuning MLFF on more experimental measurements, and development on better \textit{ab initio} methods for simulating transport properties of organic liquids. Moreover, the acceleration of the model further facilitates its industrial applications. We hope this study can pave the way for designing better thermal conducting or insulating liquids for diverse applications by atomistic simulation.

\section{Code availability}\label{code}
The source codes, including the GEDT model, the training module and the LAMMPS interface for the MD simulations, are available via GitHub at https://github.com/bytedance/bamboo at branch feat/thermal.

\section{Author contributions}
Wei Feng: Conceptualization, Methodology, Investigation, Writing; Siyuan Liu: Conceptualization, Methodology, Investigation, Writing; Hongyi Wang: Conceptualization, Methodology, Investigation, Writing; Zhenliang Mu: Conceptualization, Methodology, Investigation; Zhichen Pu: Conceptualization, Methodology, Investigation; Xu Han: Conceptualization, Methodology; Tianze Zheng: Conceptualization, Methodology; Zhenze Yang: Investigation; Zhi Wang: Conceptualization, Methodology; Weihao Gao: Conceptualization, Methodology; Yidan Cao: Supervision; Kuang Yu: Conceptualization, Methodology, Investigation, Writing, Supervision; Sheng Gong: Conceptualization, Methodology, Investigation, Writing, Supervision; Wen Yan: Conceptualization, Methodology, Investigation, Supervision.

Wei Feng and Siyuan Liu contributed equally to this work.

\section{Competing interests}
The authors declare no competing interests.

\clearpage

\bibliographystyle{plainnat}
\bibliography{main}

@article{gong2024bamboo,
  title={A predictive machine learning force-field framework for liquid electrolyte development},
  author={Gong, Sheng and Zhang, Yumin and Mu, Zhenliang and Pu, Zhichen and Wang, Hongyi and Han, Xu and Yu, Zhiao and Chen, Mengyi and Zheng, Tianze and Wang, Zhi and others},
  journal={Nature Machine Intelligence},
  pages={1--10},
  year={2025},
  publisher={Nature Publishing Group UK London}
}

@article{ye2024differential,
  title={Differential transformer},
  author={Ye, Tianzhu and Dong, Li and Xia, Yuqing and Sun, Yutao and Zhu, Yi and Huang, Gao and Wei, Furu},
  journal={arXiv preprint arXiv:2410.05258},
  year={2024}
}

@book{laplante2018comprehensive,
  title={Comprehensive dictionary of electrical engineering},
  author={Laplante, Philip A and Cravey, Robin and Dunleavy, Lawrence P and Antonakos, James L and LeRoy, Rodney and East, Jack and Buris, Nicholas E and Conant, Christopher J and Fryda, Lawrence and Boyd, Robert William and others},
  year={2018},
  publisher={CRC Press}
}

@article{mitra2022cite01,
  title={Advances in the improvement of thermal-conductivity of phase change material-based lithium-ion battery thermal management systems: An updated review},
  author={Mitra, Abhijeet and Kumar, Rajan and Singh, Dwesh Kumar and Said, Zafar},
  journal={Journal of Energy Storage},
  volume={53},
  pages={105195},
  year={2022},
  publisher={Elsevier}
}

@article{sanker2022cite02,
  title={Phase change material based thermal management of lithium ion batteries: A review on thermal performance of various thermal conductivity enhancers},
  author={Sanker, S Babu and Baby, Rajesh},
  journal={Journal of Energy Storage},
  volume={50},
  pages={104606},
  year={2022},
  publisher={Elsevier}
}

@article{yoo2007cite03,
  title={Study of thermal conductivity of nanofluids for the application of heat transfer fluids},
  author={Yoo, Dae-Hwang and Hong, KS and Yang, Ho-Soon},
  journal={Thermochimica Acta},
  volume={455},
  number={1-2},
  pages={66--69},
  year={2007},
  publisher={Elsevier}
}

@article{franca2018cite04,
  title={Thermal conductivity of ionic liquids and ionanofluids and their feasibility as heat transfer fluids},
  author={Fran\c{c}a, Jo\~{a}o MP and Louren\c{c}o, Maria Jos\'{e} V and Murshed, SM Sohel and P\'{a}dua, Ag\'{i}lio AH and Nieto de Castro, Carlos A},
  journal={Industrial \& Engineering Chemistry Research},
  volume={57},
  number={18},
  pages={6516--6529},
  year={2018},
  publisher={ACS Publications}
}

@article{van2005cite05,
  title={Thermochemistry of ionic liquid heat-transfer fluids},
  author={Van Valkenburg, Michael E and Vaughn, Robert L and Williams, Margaret and Wilkes, John S},
  journal={Thermochimica Acta},
  volume={425},
  number={1-2},
  pages={181--188},
  year={2005},
  publisher={Elsevier}
}

@article{mohapatra2005cite06,
  title={Advances in liquid coolant technologies for electronics cooling},
  author={Mohapatra, Satish C and Loikits, Daniel},
  journal={Semiconductor Thermal Measurement and Management IEEE Twenty First Annual IEEE Symposium, 2005.},
  pages={354--360},
  year={2005},
  organization={IEEE}
}

@article{muller1997rnemd,
  title={A simple nonequilibrium molecular dynamics method for calculating the thermal conductivity},
  author={M{\"u}ller-Plathe, Florian},
  journal={The Journal of chemical physics},
  volume={106},
  number={14},
  pages={6082--6085},
  year={1997},
  publisher={American Institute of Physics}
}

@article{zhang2023dp_thermal,
  title={Thermal conductivity of water at extreme conditions},
  author={Zhang, Cunzhi and Puligheddu, Marcello and Zhang, Linfeng and Car, Roberto and Galli, Giulia},
  journal={The Journal of Physical Chemistry B},
  volume={127},
  number={31},
  pages={7011--7017},
  year={2023},
  publisher={ACS Publications}
}

@article{tisi2021dp_heat,
  title={Heat transport in liquid water from first-principles and deep neural network simulations},
  author={Tisi, Davide and Zhang, Linfeng and Bertossa, Riccardo and Wang, Han and Car, Roberto and Baroni, Stefano},
  journal={Physical Review B},
  volume={104},
  number={22},
  pages={224202},
  year={2021},
  publisher={APS}
}

@article{wang2008electrolyte01,
  title={Modeling thermal conductivity of concentrated and mixed-solvent electrolyte systems},
  author={Wang, Peiming and Anderko, Andrzej},
  journal={Industrial \& engineering chemistry research},
  volume={47},
  number={15},
  pages={5698--5709},
  year={2008},
  publisher={ACS Publications}
}

@article{takami2001electrolyte02,
  title={New thin lithium-ion batteries using a liquid electrolyte with thermal stability},
  author={Takami, Norio and Sekino, Masahiro and Ohsaki, Takahisa and Kanda, Motoya and Yamamoto, Masao},
  journal={Journal of power sources},
  volume={97},
  pages={677--680},
  year={2001},
  publisher={Elsevier}
}

@article{ye2024electrolyte03,
  title={Development of the electrolyte in lithium-ion battery: a concise review on its thermal hazards},
  author={Ye, Jia-Chi and Lai, Yen-Wen and Huang, Xin-Hao and Chang, Zhi-Xiang and Chung, Yi-Hung and Shu, Chi-Min},
  journal={Journal of Thermal Analysis and Calorimetry},
  pages={1--20},
  year={2024},
  publisher={Springer}
}

@article{chantrenne2004finite_size,
  title={Finite size effects in determination of thermal conductivities: comparing molecular dynamics results with simple models},
  author={Chantrenne, Patrice and Barrat, Jean-Louis},
  journal={J. Heat Transfer},
  volume={126},
  number={4},
  pages={577--585},
  year={2004},
  publisher={American Society of Mechanical Engineers}
}

@article{thompson2022lammps,
  title={LAMMPS-a flexible simulation tool for particle-based materials modeling at the atomic, meso, and continuum scales},
  author={Thompson, Aidan P and Aktulga, H Metin and Berger, Richard and Bolintineanu, Dan S and Brown, W Michael and Crozier, Paul S and In't Veld, Pieter J and Kohlmeyer, Axel and Moore, Stan G and Nguyen, Trung Dac and others},
  journal={Computer Physics Communications},
  volume={271},
  pages={108171},
  year={2022},
  publisher={Elsevier}
}

@article{van2005gromacs,
  title={GROMACS: fast, flexible, and free},
  author={Van Der Spoel, David and Lindahl, Erik and Hess, Berk and Groenhof, Gerrit and Mark, Alan E and Berendsen, Herman JC},
  journal={Journal of computational chemistry},
  volume={26},
  number={16},
  pages={1701--1718},
  year={2005},
  publisher={Wiley Online Library}
}

@article{paszke2019pytorch,
  title={Pytorch: An imperative style, high-performance deep learning library},
  author={Paszke, Adam and Gross, Sam and Massa, Francisco and Lerer, Adam and Bradbury, James and Chanan, Gregory and Killeen, Trevor and Lin, Zeming and Gimelshein, Natalia and Antiga, Luca and others},
  journal={Advances in neural information processing systems},
  volume={32},
  year={2019}
}

@book{yaws2009tc_eib,
  title={Thermal Conductivity of Liquid--Organic Compounds},
  author={Yaws, Carl L},
  booktitle={Transport Properties of Chemicals and Hydrocarbons},
  pages={299--395},
  year={2009},
  publisher={Elsevier}
}

@book{yaws1995handbook2,
  title={Handbook of thermal conductivity, volume 2: organic compounds C5 To C7},
  author={Yaws, Carl L},
  year={1995},
  publisher={Elsevier}
}

@book{yaws1997handbook1,
  title={Handbook of thermal conductivity, Volume 1: Organic Compounds C1 to C4},
  author={Yaws, CL},
  year={1997},
  publisher={Elsevier}
}

@misc{tc_ef_293,
  key = {CAMEO Chemicals NOAA},
  url = {https://cameochemicals.noaa.gov/chris/EFM.pdf},
}

@misc{tc_mo,
  key = {Material Data Sheet},
  url = {https://www.matweb.com/search/datasheet.aspx?matguid=f2c9a5d8608e4f5aac1d570d37a54ffc&ckck=1},
}

@misc{tc_eo,
  key = {Thermtest Asia Material Database},
  url = {https://thermtestasia.cn/material-database},
}

@article{chen2002density_mo,
  title={Density and Refractive Index at 298.15 K and Vapor- Liquid Equilibria at 101.3 kPa for Four Binary Systems of Methanol, n-Propanol, n-Butanol, or Isobutanol with N-Methylpiperazine},
  author={Chen, Shuda and Lei, Qunfang and Fang, Wenjun},
  journal={Journal of Chemical \& Engineering Data},
  volume={47},
  number={4},
  pages={811--815},
  year={2002},
  publisher={ACS Publications}
}

@article{ortega1982density_eo,
  title={Densities and refractive indices of pure alcohols as a function of temperature},
  author={Ortega, Juan},
  journal={Journal of Chemical and Engineering Data},
  volume={27},
  number={3},
  pages={312--317},
  year={1982},
  publisher={ACS Publications}
}

@article{erastova2017density_tbuo,
  title={Understanding surface interactions in aqueous miscible organic solvent treated layered double hydroxides},
  author={Erastova, Valentina and Degiacomi, Matteo T and O'Hare, Dermot and Greenwell, H Chris},
  journal={RSC advances},
  volume={7},
  number={9},
  pages={5076--5083},
  year={2017},
  publisher={Royal Society of Chemistry}
}

@article{ohta1980density_ef,
  title={Thermodynamic properties of four ester-hydrocarbon mixtures},
  author={Ohta, Tatsuhiko and Nagata, Isamu},
  journal={Journal of Chemical and Engineering Data},
  volume={25},
  number={3},
  pages={283--286},
  year={1980},
  publisher={ACS Publications}
}

@article{sastry2013density_ep,
  title={Excess molar volumes, excess isentropic compressibilities, excess viscosities, relative permittivity and molar polarization deviations for methyl acetate+, ethyl acetate+, butyl acetate+, isoamyl acetate+, methyl propionate+, ethyl propionate+, ethyl butyrate+, methyl methacrylate+, ethyl methacrylate+, and butyl methacrylate+ cyclohexane at T= 298.15 and 303.15 K},
  author={Sastry, Nandhibatla V and Patel, Sunil R and Soni, Saurabh S},
  journal={Journal of Molecular Liquids},
  volume={183},
  pages={102--112},
  year={2013},
  publisher={Elsevier}
}

@article{wang2020density_ipa,
  title={Material properties of porous asphalt pavement cold patch mixtures with different solvents},
  author={Wang, Xiang and Chen, Xueqin and Dong, Qiao and Jahanzaib, Ahmad},
  journal={Journal of Materials in Civil Engineering},
  volume={32},
  number={10},
  pages={06020015},
  year={2020},
  publisher={American Society of Civil Engineers}
}

@article{byrne2018density_eib,
  title={A methodical selection process for the development of ketones and esters as bio-based replacements for traditional hydrocarbon solvents},
  author={Byrne, Fergal P and Forier, Bart and Bossaert, Greet and Hoebers, Charly and Farmer, Thomas J and Hunt, Andrew J},
  journal={Green Chemistry},
  volume={20},
  number={17},
  pages={4003--4011},
  year={2018},
  publisher={Royal Society of Chemistry}
}

@article{gardas2007density_pra_882,
  title={PVT Property Measurements for Some Aliphatic Esters from (298 to 393) K and up to 35 MPa},
  author={Gardas, Ramesh L and Johnson, Irudayaraj and Vaz, David MD and Fonseca, Isabel MA and Ferreira, Abel GM},
  journal={Journal of Chemical \& Engineering Data},
  volume={52},
  number={3},
  pages={737--751},
  year={2007},
  publisher={ACS Publications}
}

@misc{fao2024_ma_93,
  author       = {Food and Agriculture Organization},
  title        = {JECFA Flavonoids - Details},
  url = {https://www.fao.org/food/food-safety-quality/scientific-advice/jecfa/jecfa-flav/details/en/c/245/}
}

@article{sykioti2013methanol_tc,
  title={Reference Correlation of the Thermal Conductivity of Methanol from the Triple Point to 660 K and up to 245 MPa},
  author={Sykioti, EA and Assael, Marc J and Huber, Marcia L and Perkins, Richard A},
  journal={Journal of Physical and Chemical Reference Data},
  volume={42},
  number={4},
  year={2013},
  publisher={AIP Publishing}
}

@article{perdew1996pbe_gga,
  title={Generalized gradient approximation made simple},
  author={Perdew, John P and Burke, Kieron and Ernzerhof, Matthias},
  journal={Physical review letters},
  volume={77},
  number={18},
  pages={3865},
  year={1996},
  publisher={APS}
}

@article{sun2015scan_meta_gga,
  title={Strongly constrained and appropriately normed semilocal density functional},
  author={Sun, Jianwei and Ruzsinszky, Adrienn and Perdew, John P},
  journal={Physical review letters},
  volume={115},
  number={3},
  pages={036402},
  year={2015},
  publisher={APS}
}

@article{raghavachari1989ccsdt,
  title={A fifth-order perturbation comparison of electron correlation theories},
  author={Raghavachari, Krishnan and Trucks, Gary W and Pople, John A and Head-Gordon, Martin},
  journal={Chemical Physics Letters},
  volume={157},
  number={6},
  pages={479--483},
  year={1989},
  publisher={Elsevier}
}

@article{sellan2010size_effect,
  title={Size effects in molecular dynamics thermal conductivity predictions},
  author={Sellan, Daniel P and Landry, Eric S and Turney, JE and McGaughey, Alan JH and Amon, Cristina H},
  journal={Physical Review B—Condensed Matter and Materials Physics},
  volume={81},
  number={21},
  pages={214305},
  year={2010},
  publisher={APS}
}

@article{zhang2005water_classical_force_field,
  title={Thermal conductivities of molecular liquids by reverse nonequilibrium molecular dynamics},
  author={Zhang, Meimei and Lussetti, Enrico and de Souza, Lu{\'\i}s ES and M{\"u}ller-Plathe, Florian},
  journal={The Journal of Physical Chemistry B},
  volume={109},
  number={31},
  pages={15060--15067},
  year={2005},
  publisher={ACS Publications}
}

@article{becke1997gga_2,
  title={Density-functional thermochemistry. V. Systematic optimization of exchange-correlation functionals},
  author={Becke, Axel D},
  journal={The Journal of chemical physics},
  volume={107},
  number={20},
  pages={8554--8560},
  year={1997},
  publisher={American Institute of Physics}
}

@article{becke1988gga_1,
  title={Density-functional exchange-energy approximation with correct asymptotic behavior},
  author={Becke, Axel D},
  journal={Physical review A},
  volume={38},
  number={6},
  pages={3098},
  year={1988},
  publisher={APS}
}

@article{purvis1982ccsd_t,
  title={A full coupled-cluster singles and doubles model: The inclusion of disconnected triples},
  author={Purvis, George D and Bartlett, Rodney J},
  journal={The Journal of Chemical Physics},
  volume={76},
  number={4},
  pages={1910--1918},
  year={1982},
  publisher={AIP Publishing}
}

@article{cheng2020size_eq,
  title={Computing the heat conductivity of fluids from density fluctuations},
  author={Cheng, Bingqing and Frenkel, Daan},
  journal={Physical Review Letters},
  volume={125},
  number={13},
  pages={130602},
  year={2020},
  publisher={APS}
}

@article{unke2021machine,
  title={Machine learning force fields},
  author={Unke, Oliver T and Chmiela, Stefan and Sauceda, Huziel E and Gastegger, Michael and Poltavsky, Igor and Schu\"{u}tt, Kristof T and Tkatchenko, Alexandre and M\"{u}ller, Klaus-Robert},
  journal={Chem. Rev.},
  volume={121},
  number={16},
  pages={10142--10186},
  year={2021},
  publisher={ACS Publications}
}

@book{frenkel2023understanding,
  title={Understanding molecular simulation: from algorithms to applications},
  author={Frenkel, Daan and Smit, Berend},
  year={2023},
  publisher={Elsevier}
}

@article{2015_Schröder,
author = {Heiner Schröder and Anne Creon and Tobias Schwabe},
title = {Reformulation of the D3(Becke–Johnson) Dispersion Correction without Resorting to Higher than C6 Dispersion Coefficients},
journal = {J. Chem. Theory Comput.},
year = {2015},
volume = {11},
publisher = {American Chemical Society (ACS)},
month = {6},
number = {7},
pages = {3163--3170},
doi = {10.1021/acs.jctc.5b00400}}

@article{TORRESSANCHEZ2016224,
title = {Geometric derivation of the microscopic stress: A covariant central force decomposition},
journal = {J. Mech. Phys. Solids},
volume = {93},
pages = {224-239},
year = {2016},
note = {Special Issue in honor of Michael Ortiz},
issn = {0022-5096},
doi = {https://doi.org/10.1016/j.jmps.2016.03.006},
author = {Alejandro Torres-Sánchez and Juan M. Vanegas and Marino Arroyo}}

@article{vaswani2017attention,
  title={Attention is all you need},
  author={Vaswani, Ashish and Shazeer, Noam and Parmar, Niki and Uszkoreit, Jakob and Jones, Llion and Gomez, Aidan N and Kaiser, {\L}ukasz and Polosukhin, Illia},
  journal={Advances in neural information processing systems},
  volume={30},
  year={2017}
}

@article{rbf,
author = {Unke, Oliver and Meuwly, Markus},
year = {2019},
month = {05},
pages = {},
title = {PhysNet: A Neural Network for Predicting Energies, Forces, Dipole Moments and Partial Charges},
volume = {15},
journal = {J. Chem. Theory Comput.},
doi = {10.1021/acs.jctc.9b00181}
}

@article{silu,
  title={Sigmoid-weighted linear units for neural network function approximation in reinforcement learning},
  author={Elfwing, Stefan and Uchibe, Eiji and Doya, Kenji},
  journal={Neural networks},
  volume={107},
  pages={3--11},
  year={2018},
  publisher={Elsevier}
}

@article{b3lyp,
    author = {Becke, Axel D.},
    title = "{Density‐functional thermochemistry. III. The role of exact exchange}",
    journal = {J. Chem. Phys.},
    volume = {98},
    number = {7},
    pages = {5648-5652},
    year = {1993},
    month = {04},
    issn = {0021-9606},
    doi = {10.1063/1.464913},
}

@Article{svpd,
author ="Hellweg, Arnim and Rappoport, Dmitrij",
title  ="Development of new auxiliary basis functions of the Karlsruhe segmented contracted basis sets including diffuse basis functions (def2-SVPD{,} def2-TZVPPD{,} and def2-QVPPD) for RI-MP2 and RI-CC calculations",
journal  ="Phys. Chem. Chem. Phys.",
year  ="2015",
volume  ="17",
issue  ="2",
pages  ="1010-1017",
publisher  ="The Royal Society of Chemistry",
doi  ="10.1039/C4CP04286G",
}

@article{m3gnet,
author = {Chen, Chi and Ong, Shyue},
year = {2022},
month = {11},
pages = {718-728},
title = {A universal graph deep learning interatomic potential for the periodic table},
volume = {2},
journal = {Nat. Comput. Sci.},
doi = {10.1038/s43588-022-00349-3}
}

@article{oplsaa,
author = {Jorgensen, William L. and Maxwell, David S. and Tirado-Rives, Julian},
title = {Development and Testing of the OPLS All-Atom Force Field on Conformational Energetics and Properties of Organic Liquids},
journal = {Journal of the American Chemical Society},
volume = {118},
number = {45},
pages = {11225-11236},
year = {1996},
}

@software{code,
  author       = {muzhenliang},
  title        = {muzhenliang/bamboo: v0.1},
  month        = jan,
  year         = 2025,
  publisher    = {Zenodo},
  version      = {paper-v1},
  doi          = {10.5281/zenodo.14603020},
  url          = {https://doi.org/10.5281/zenodo.14603020},
  swhid        = {swh:1:dir:c7fd788d3c3e966667f37919a4093b5f83601e5d
                   ;origin=https://doi.org/10.5281/zenodo.14603019;vi
                   sit=swh:1:snp:21e461a910f2162e8c7a413510a7b8c97f17
                   6d72;anchor=swh:1:rel:00c4b9859710b059c81c7a4c6653
                   7d1daa577458;path=muzhenliang-bamboo-21ce529
                  },
}

@article{chgnet,
title = {CHGNet as a pretrained universal neural network potential for charge-informed atomistic modelling},
author = {Deng, Bowen and Zhong, Peichen and Jun, KyuJung and Riebesell, Janosh and Han, Kevin and Bartel, Christopher J. and Ceder, Gerbrand},
doi = {10.1038/s42256-023-00716-3},
journal = {Nat. Mach. Intell.},
number = 9,
volume = 5,
place = {United States},
year = {2023},
month = {9}
}

@article{genome,
author = {Merchant, Amil and Batzner, Simon and Schoenholz, Samuel and Aykol, Muratahan and Cheon, Gowoon and Cubuk, Ekin},
year = {2023},
month = {11},
pages = {1-6},
title = {Scaling deep learning for materials discovery},
volume = {624},
journal = {Nature},
doi = {10.1038/s41586-023-06735-9}
}

@misc{dpa2,
      title={DPA-2: Towards a universal large atomic model for molecular and material simulation}, 
      author={Duo Zhang and Xinzijian Liu and Xiangyu Zhang and Chengqian Zhang and Chun Cai and Hangrui Bi and Yiming Du and Xuejian Qin and Jiameng Huang and Bowen Li and Yifan Shan and Jinzhe Zeng and Yuzhi Zhang and Siyuan Liu and Yifan Li and Junhan Chang and Xinyan Wang and Shuo Zhou and Jianchuan Liu and Xiaoshan Luo and Zhenyu Wang and Wanrun Jiang and Jing Wu and Yudi Yang and Jiyuan Yang and Manyi Yang and Fu-Qiang Gong and Linshuang Zhang and Mengchao Shi and Fu-Zhi Dai and Darrin M. York and Shi Liu and Tong Zhu and Zhicheng Zhong and Jian Lv and Jun Cheng and Weile Jia and Mohan Chen and Guolin Ke and Weinan E and Linfeng Zhang and Han Wang},
      year={2023},
      eprint={2312.15492},
      archivePrefix={arXiv},
      primaryClass={physics.chem-ph}
}

@misc{macemp,
      title={A foundation model for atomistic materials chemistry}, 
      author={Ilyes Batatia and Philipp Benner and Yuan Chiang and Alin M. Elena and Dávid P. Kovács and Janosh Riebesell and Xavier R. Advincula and Mark Asta and Matthew Avaylon and William J. Baldwin and Fabian Berger and Noam Bernstein and Arghya Bhowmik and Samuel M. Blau and Vlad Cărare and James P. Darby and Sandip De and Flaviano Della Pia and Volker L. Deringer and Rokas Elijošius and Zakariya El-Machachi and Fabio Falcioni and Edvin Fako and Andrea C. Ferrari and Annalena Genreith-Schriever and Janine George and Rhys E. A. Goodall and Clare P. Grey and Petr Grigorev and Shuang Han and Will Handley and Hendrik H. Heenen and Kersti Hermansson and Christian Holm and Jad Jaafar and Stephan Hofmann and Konstantin S. Jakob and Hyunwook Jung and Venkat Kapil and Aaron D. Kaplan and Nima Karimitari and James R. Kermode and Namu Kroupa and Jolla Kullgren and Matthew C. Kuner and Domantas Kuryla and Guoda Liepuoniute and Johannes T. Margraf and Ioan-Bogdan Magdău and Angelos Michaelides and J. Harry Moore and Aakash A. Naik and Samuel P. Niblett and Sam Walton Norwood and Niamh O'Neill and Christoph Ortner and Kristin A. Persson and Karsten Reuter and Andrew S. Rosen and Lars L. Schaaf and Christoph Schran and Benjamin X. Shi and Eric Sivonxay and Tamás K. Stenczel and Viktor Svahn and Christopher Sutton and Thomas D. Swinburne and Jules Tilly and Cas van der Oord and Eszter Varga-Umbrich and Tejs Vegge and Martin Vondrák and Yangshuai Wang and William C. Witt and Fabian Zills and Gábor Csányi},
      year={2024},
      eprint={2401.00096},
      archivePrefix={arXiv},
      primaryClass={physics.chem-ph}
}

@article{dmff,
author = {Wang, Xinyan and Li, Jichen and Yang, Lan and Chen, Feiyang and Wang, Yingze and Chang, Junhan and Chen, Junmin and Feng, Wei and Zhang, Linfeng and Yu, Kuang},
title = {DMFF: An Open-Source Automatic Differentiable Platform for Molecular Force Field Development and Molecular Dynamics Simulation},
journal = {J. Chem. Theory Comput.},
volume = {19},
number = {17},
pages = {5897-5909},
year = {2023},
doi = {10.1021/acs.jctc.2c01297},
note ={PMID: 37589304},
}

@book{crchandbook,
  title={CRC Handbook of Chemistry and Physics},
  author={Haynes, W.M.},
  isbn={9781439855126},
  series={CRC Handbook of Chemistry and Physics},
  year={2011},
  publisher={CRC Press}
}

@article{gpu4pyscf,
author = {Wu, Xiaojie and Sun, Qiming and Pu, Zhichen and Zheng, Tianze and Ma, Wenzhi and Yan, Wen and Xia, Yu and Wu, Zhengxiao and Huo, Mian and Li, Xiang and Ren, Weiluo and Gong, Sheng and Zhang, Yumin and Gao, Weihao},
title = {Enhancing GPU-Acceleration in the Python-Based Simulations of Chemistry Frameworks},
journal = {WIREs Computational Molecular Science},
volume = {15},
number = {2},
pages = {e70008},
doi = {https://doi.org/10.1002/wcms.70008},
url = {https://wires.onlinelibrary.wiley.com/doi/abs/10.1002/wcms.70008},
eprint = {https://wires.onlinelibrary.wiley.com/doi/pdf/10.1002/wcms.70008},
note = {e70008 CMS-1146.R2},
year = {2025}

}

@article{maceoff23,
  title={MACE-OFF23: Transferable machine learning force fields for organic molecules},
  author={Kov{\'a}cs, D{\'a}vid P{\'e}ter and Moore, J Harry and Browning, Nicholas J and Batatia, Ilyes and Horton, Joshua T and Kapil, Venkat and Magd{\u{a}}u, Ioan-Bogdan and Cole, Daniel J and Cs{\'a}nyi, G{\'a}bor},
  journal={arXiv preprint arXiv:2312.15211},
  year={2023}
}

@article{galli,
  title={Atomistic simulations of the thermal conductivity of liquids},
  author={Puligheddu, Marcello and Galli, Giulia},
  journal={Physical Review Materials},
  volume={4},
  number={5},
  pages={053801},
  year={2020},
  publisher={APS}
}

@article{smith2019approaching,
  title={Approaching coupled cluster accuracy with a general-purpose neural network potential through transfer learning},
  author={Smith, Justin S and Nebgen, Benjamin T and Zubatyuk, Roman and Lubbers, Nicholas and Devereux, Christian and Barros, Kipton and Tretiak, Sergei and Isayev, Olexandr and Roitberg, Adrian E},
  journal={Nature communications},
  volume={10},
  number={1},
  pages={2903},
  year={2019},
  publisher={Nature Publishing Group UK London}
}

@article{han2025refining,
  title={Refining potential energy surface through dynamical properties via differentiable molecular simulation},
  author={Han, Bin and Yu, Kuang},
  journal={Nature Communications},
  volume={16},
  number={1},
  pages={816},
  year={2025},
  publisher={Nature Publishing Group UK London}
}

@article{doublehybrid,
  title={Double-hybrid density functional theory for excited electronic states of molecules},
  author={Grimme, Stefan and Neese, Frank},
  journal={The Journal of chemical physics},
  volume={127},
  number={15},
  year={2007},
  publisher={AIP Publishing}
}

@article{cheng2020computing,
  title={Computing the heat conductivity of fluids from density fluctuations},
  author={Cheng, Bingqing and Frenkel, Daan},
  journal={Physical Review Letters},
  volume={125},
  number={13},
  pages={130602},
  year={2020},
  publisher={APS}
}

@article{mlffheatflux,
    author = {Langer, Marcel F. and Frank, J. Thorben and Knoop, Florian},
    title = {Stress and heat flux via automatic differentiation},
    journal = {The Journal of Chemical Physics},
    volume = {159},
    number = {17},
    pages = {174105},
    year = {2023},
    month = {11},
    issn = {0021-9606},
    doi = {10.1063/5.0155760},
    url = {https://doi.org/10.1063/5.0155760},
    eprint = {https://pubs.aip.org/aip/jcp/article-pdf/doi/10.1063/5.0155760/18196139/174105\_1\_5.0155760.pdf},
}

@article{watermodel,
  title={Melting points of water models: Current situation},
  author={Blazquez, Samuel and Vega, Carlos},
  journal={The Journal of Chemical Physics},
  volume={156},
  number={21},
  year={2022},
  publisher={AIP Publishing}
}

@article{xu2023nqe_water,
  title={Accurate prediction of heat conductivity of water by a neuroevolution potential},
  author={Xu, Ke and Hao, Yongchao and Liang, Ting and Ying, Penghua and Xu, Jianbin and Wu, Jianyang and Fan, Zheyong},
  journal={The Journal of Chemical Physics},
  volume={158},
  number={20},
  year={2023},
  publisher={AIP Publishing}
}

@article{xu2024nqe_density,
  title={NEP-MB-pol: A unified machine-learned framework for fast and accurate prediction of water's thermodynamic and transport properties},
  author={Xu, Ke and Liang, Ting and Xu, Nan and Ying, Penghua and Chen, Shunda and Wei, Ning and Xu, Jianbin and Fan, Zheyong},
  journal={arXiv preprint arXiv:2411.09631},
  year={2024}
}

@article{luo2020nqe_ice,
  title={Capturing the nuclear quantum effects in molecular dynamics for lattice thermal conductivity calculations: Using ice as example},
  author={Luo, Ripeng and Yu, Kuang},
  journal={The Journal of Chemical Physics},
  volume={153},
  number={19},
  year={2020},
  publisher={AIP Publishing}
}

@inproceedings{tillet2019triton,
  title={Triton: an intermediate language and compiler for tiled neural network computations},
  author={Tillet, Philippe and Kung, Hsiang-Tsung and Cox, David},
  booktitle={Proceedings of the 3rd ACM SIGPLAN International Workshop on Machine Learning and Programming Languages},
  pages={10--19},
  year={2019}
}

@article{phyneo,
	title = {{PhyNEO}: A Neural-Network-Enhanced Physics-Driven Force Field Development Workflow for Bulk Organic Molecule and Polymer Simulations},
    author = {Chen, Junmin and Yu, Kuang},
    journal = {Journal of Chemical Theory and Computation},	
    volume = {20},
    number = {1},
	pages = {253--265},
    year={2024}
}

@article{tu_enhancing_2025,
	title = {Enhancing Thermal Conductivity Computation of Polymers via Machine Learning Techniques},
    author = {Tu, Chengyang and Li, Xin and Chen, Junmin and Sun, Bo and Yu, Kuang},
    journal = {The Journal of Physical Chemistry B},
	volume = {129},
    number = {33},
	pages = {8593--8602},
	year={2025}
}

@article{feng_screening_2025,
	title = {Screening and Design of Aqueous Zinc Battery Electrolytes Based on the Multimodal Optimization of Molecular Simulation},
    author = {Feng, Wei and Zhang, Luyan and Cheng, Yaobo and Wu, Jin and Wei, Chunguang and Zhang, Junwei and Yu, Kuang},
    journal = {The Journal of Physical Chemistry Letters},	
    volume = {16},
	number = {13},
	pages = {3326--3335},
    year={2025}
}

@article{zhuang1,
	title = {Exceptional piezoelectricity, high thermal conductivity and stiffness and promising photocatalysis in two-dimensional MoSi2N4 family confirmed by first-principles},
    author = {Bohayra, Mortazavi and Brahmanandam, Javvaji and Fazel, Shojaei and Timon, Rabczuk and Alexander V. Shapeev and Xiaoying, Zhuang},
    journal = {Nano Energy},	
    volume = {82},
	pages = {105716},
    year={2021}
}

@article{zhuang2,
	title = {Accelerating first-principles estimation of thermal conductivity by machine-learning interatomic potentials: A MTP/ShengBTE solution},
    author = {Bohayra, Mortazavi and Evgeny, V. Podryabinkin and Ivan, S. Novikov and Timon, Rabczuk and Xiaoying, Zhuang and Alexander, V. Shapeev},
    journal = {Computer Physics Communications},	
    volume = {258},
	pages = {107583},
    year={2021}
}

@article{thermalcond,
	title = {Measurement of the thermal conductivity of five aliphatic esters in the liquid phase},
    author = {Fenhong, Song and Dapeng, Ju and Jing, Fan and Xiaopo, Wang and Gang, Wang},
    journal = {The Journal of Chemical Thermodynamics},	
    volume = {138},
	pages = {140-146},
    year={2019}
}

@article{density1,
	title = {Dicyclic Hydrocarbons. IV. Synthesis and Physical Properties of {$\alpha$}, {$\alpha$}- and {$\alpha$},{$\iota$}-Diphenyl- and Dicyclohexyl- Pentanes and Hexanes1},
    author = {Kasper T. Serijan and Paul H. Wise},
    journal = {Journal of the American Chemical Society},	
    volume = {73, 11},
	pages = {5191-5193},
    year={1951}
}

\clearpage

\beginappendix
\renewcommand{\thefigure}{S\arabic{figure}}
\setcounter{figure}{0}
\renewcommand{\thetable}{S\arabic{table}}
\setcounter{table}{0}
\renewcommand{\theequation}{S\arabic{equation}}
\setcounter{equation}{0}

\section*{Supplementary Information}\label{title:si}

\section*{Accelerated Machine Learning Force Field for Predicting Thermal Conductivity of Organic Liquids}

\section{Graph Equivariant Differential Transformer (GEDT)}\label{subsec:si_gedt}

\begin{figure}
    \centering
    \includegraphics[width=\linewidth]{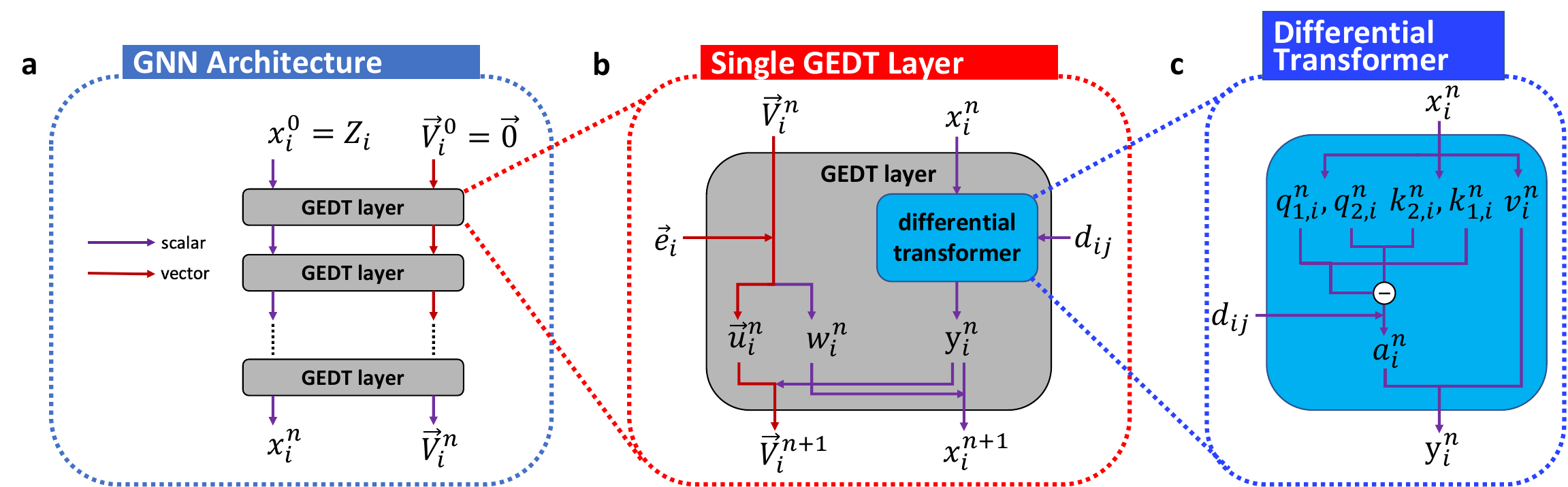}
    \caption{Schematic of the \textbf{a}: GNN, \textbf{b}: GEDT layer inside the GNN, and \textbf{c}: differential transformer inside the GEDT layer.}
    \label{fig:si_bamboo}
\end{figure}

\begin{figure}
    \centering
    \includegraphics[width=1\linewidth]{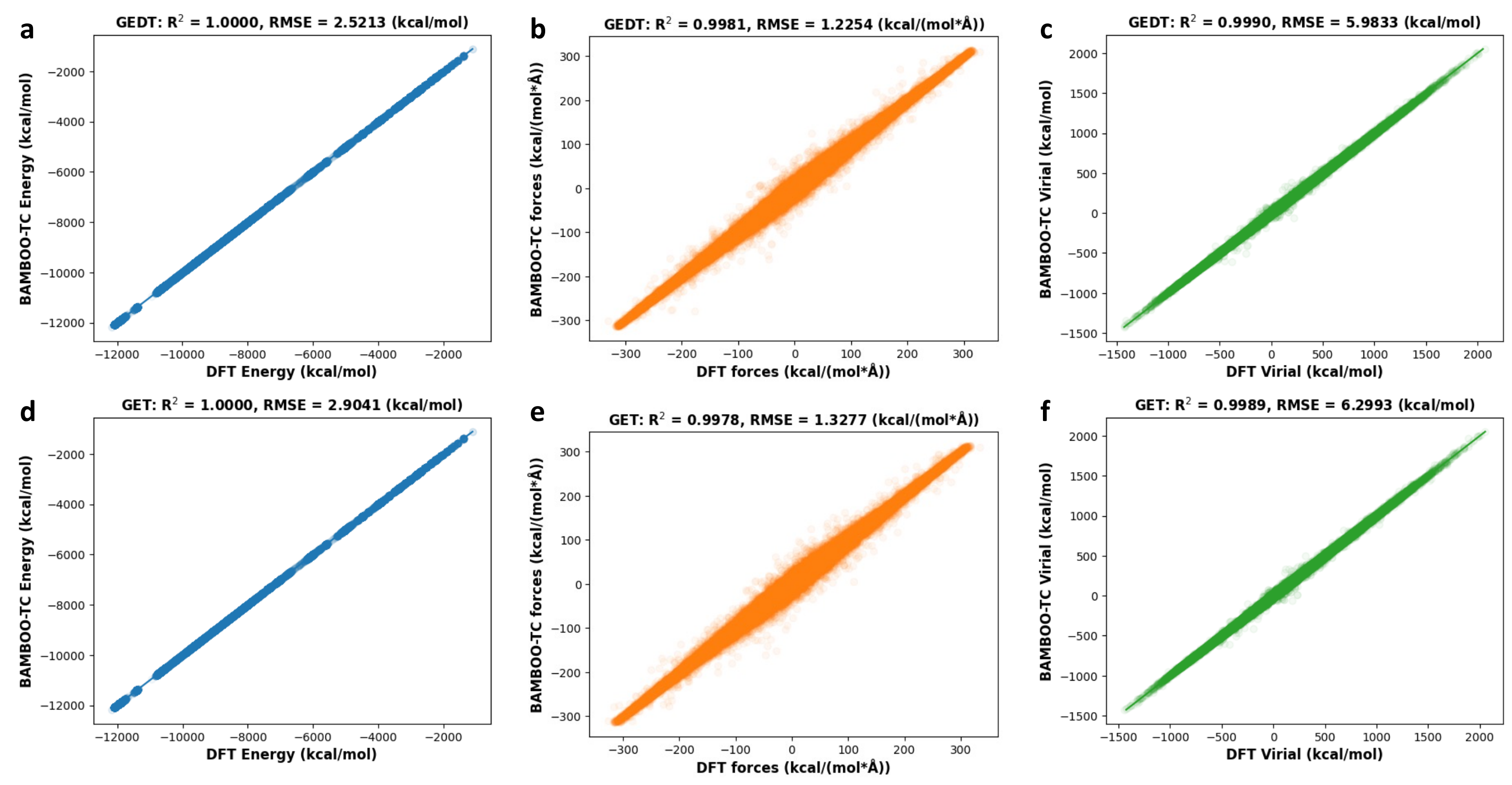}
    \caption{Fitting performance of BAMBOO-TC with GEDT and GET layers. GEDT layers fit the DFT PES better than GET layers, demonstrating higher R$^2$ scores and lower root mean square error (RMSE). \textbf{a, b, c}: BAMBOO-TC predicted energy, forces, and virial predictions with GEDT layers comparing with DFT labels in validation set. \textbf{d, e, f}: BAMBOO-TC predicted energy, forces, and virial predictions with GET layers comparing with DFT labels in validation set.}
    \label{fig:bamboo_dft}
\end{figure}

The architecture of GEDT is inspired by the Graph Equivariant Transformer (GET) model~\cite{gong2024bamboo}. The main innovation of GEDT over GET is the introduction of differential transformer~\cite{ye2024differential} illustrated in Figure~\ref{fig:si_bamboo}c, which amplifies attention to the important interatomic interactions and cancels noise widely presented in the original transformer architecture~\cite{ye2024differential}. The differential attention mechanism is applied exclusively to the atomic vector representations, thereby preserving equivariant operations in alignment with the GET introduced in BAMBOO~\cite{gong2024bamboo}. As a result, GEDT learns the DFT labels better than GET with lower validation error as shown in Figure~\ref{fig:bamboo_dft}.

Details of the GEDT model are as follows: given the atomic structures, we begin by initializing the node scalar representation $x_{i}^{0}$ and the node vector representations $\Vec{V}_{i}^{0}$ for atom $i$ according to its atom type $z_{i}$. Additionally, we set up the edge scalar representation $d_{ij}$ and the edge vector representation $\Vec{e}_{ij}$ based on the relative vector $\Vec{r}_{ij}$ between atom $i$ and atom $j$, within a specified cutoff radius $r_{\text{cut}}$, which is set as 5 \AA\ in this study:

\begin{equation}
x_{i}^{0}= \text{embeddingMLP} (z_{i}),
\label{node scalar representations initialization}
\end{equation}
\begin{equation}
\Vec{V}_{i}^{0}= \Vec{0},
\label{node vector representations initialization}
\end{equation}
\begin{equation}
\hat{r}_{ij}=\frac{\vec{r}_{ij}}{||r_{ij}||},
\end{equation}
\begin{equation}
d_{ij}= \text{SiLU} (W^{d}\text{RBF} (||r_{ij}||)),
\label{edge scalar representations initialization}
\end{equation}
\begin{equation}
\Vec{e}_{ij}=\text{SiLU} (W^{e}d_{ij})^{T}\otimes\hat{r}_{ij}.
\label{edge vector representations initialization}
\end{equation} 

In this context, ``embeddingMLP" denotes a multi-layer perceptron (MLP) that converts discrete atom types into continuous embeddings, which are implemented using the \textsc{nn.Embedding} class in PyTorch~\cite{paszke2019pytorch}. The term $||\cdot||$ represents the L2 norm of a vector, while RBF refers to the radial distribution functions-based expansion as outlined in Ref.~\cite{rbf}, which transforms distances into high-dimensional representations. SiLU indicates the Sigmoid-Weighted Linear Units~\cite{silu} activation function. In equations~\eqref{edge scalar representations initialization} and \eqref{edge vector representations initialization}, $W^{d}$ and $W^{e}$ denote the weight matrices employed to produce $d_{ij}$ and $\Vec{e}_{ij}$, respectively. We then input $x_{i}^{0}$ and $\Vec{V}_{i}^{0}$ into the GEDT layers for iterative updates (Figure \ref{fig:si_bamboo}a). In the n$^{th}$ layer, we first input $x_{i}^{n}$, $||r_{ij}||$, and $d_{ij}$ into the differential transformer as illustrated in Figure~\ref{fig:si_bamboo}c:

\begin{equation}
q_{1,i}^{n}=W^{q1,n}x^{n}_{i}+b^{q1,n},
\end{equation}
\begin{equation}
k_{1,i}^{n}=W^{k1,n}x^{n}_{i}+b^{k1,n},
\end{equation}
\begin{equation}
q_{2,i}^{n}=W^{q2,n}x^{n}_{i}+b^{q2,n},
\end{equation}
\begin{equation}
k_{2,i}^{n}=W^{k2,n}x^{n}_{i}+b^{k2,n},
\end{equation}
\begin{equation}
v_{i}^{n}=W^{v,n}x^{n}_{i}+b^{v,n}.
\label{qkv}
\end{equation}

\begin{figure}
    \centering
    \includegraphics[width=0.98\linewidth]{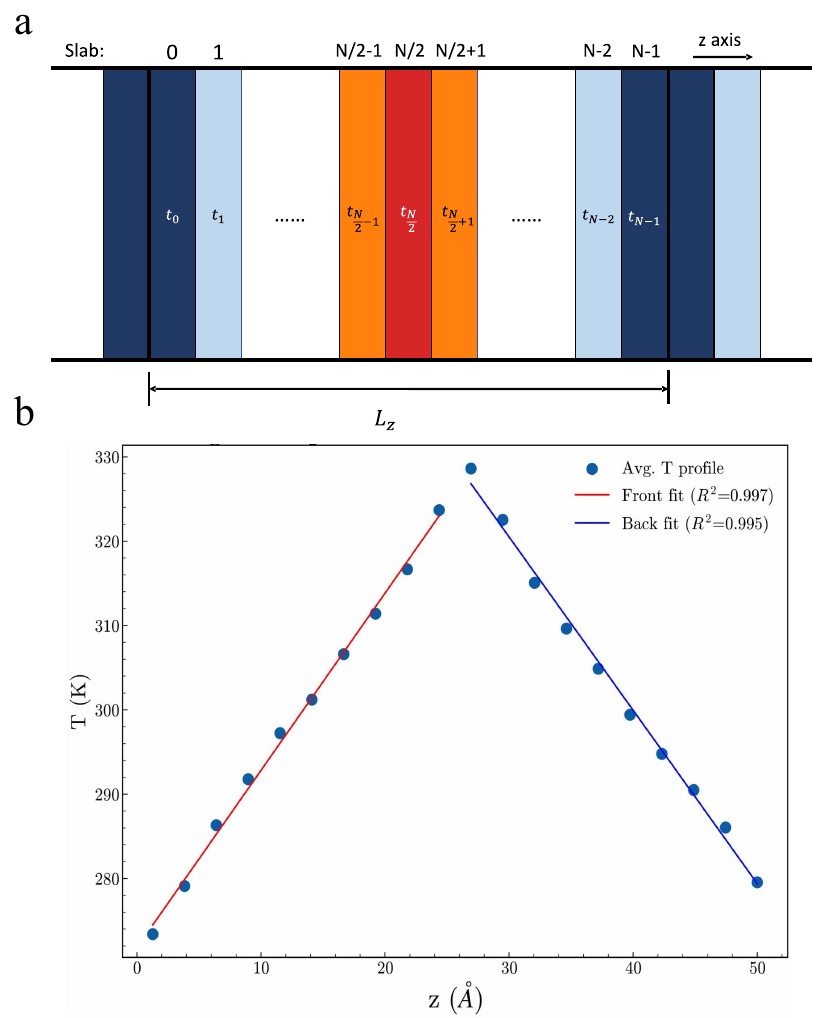}
    \caption{\textbf{a}: A demonstration of dividing the box into slabs along z-axis. The center slab N/2 is created as the hottest slab while the both ends of the box are created as the coolest slabs in this periodic system. \textbf{b}: The temperature profile of the upper and the lower half of the box in the simulation of tert-butanol.}
    \label{fig:si_thermal}
\end{figure}

Here in the differential transformer block, $q$, $k$ and $v$ represent the queries, keys, and values of the transformer. The $q$ and $k$ are divided into two vectors, respectively, for computation of the final attention score as difference between two attention functions in equation~\ref{attention}. This idea is inspired by the application of differential attention in the field of language model~\cite{ye2024differential} to eliminate attention noises, which is rooted in the concept of differential amplifiers~\cite{laplante2018comprehensive} where the difference between two signals is used as output to null out the common-mode noise of the input.

\begin{equation}
a_{ij}^{n}=\frac{1}{2}(\text{SiLU}(\langle q_{1,i}^{n}, k_{1,j}^{n}\rangle)-\text{SiLU}(\langle q_{2,i}^{n}, k_{2,j}^{n}\rangle))(\cos{\frac{\pi||r_{ij}||}{r_{\text{cut}}}+1}),
\label{attention}
\end{equation}

After the computation of differential attention, we derive the intermediate atom scalar representation $y_{i}^{n}$ on the scalar side. Zoom back to the GEDT layer, we create an intermediate vector representation $\Vec{u}_{i}^{n}$ to integrate the edge vector representation $\Vec{e}_{ij}$ on the vector side, allowing us to aggregate the contributions from neighboring atoms into the vector representation.

\begin{equation}
y_{i}^{n}=\sum_{j}a_{ji}^{n}(v_{j}^{n}\odot d_{ij}),
\label{y}
\end{equation}
\begin{equation}
\vec u_{i}^{n}=\sum_{j} v_{j}^{n} * \Vec{e}_{ij}.
\label{u}
\end{equation}

In conjunction with $\Vec{e}_{ij}$, the atom vector representation $\Vec{V}_{i}^{n}$ is transformed into an intermediate vector representation $\Vec{U}_{i}^{n}$ and an additional scalar representation $W_{i}^{n}$:

\begin{equation}
\vec U_{i}^{1,n}=W^{U^{1},n}\vec V_{i}^{n},
\end{equation}
\begin{equation}
\vec U_{i}^{2,n}=W^{U^{2},n}\vec V_{i}^{n},
\end{equation}
\begin{equation}
\vec U_{i}^{3,n}=W^{U^{3},n}\vec V_{i}^{n}.
\end{equation}

which, from vector to scalar is performed using the inner product operation to ensure rotational invariance. 

Next, we compute the inner product $\langle \cdot, \cdot \rangle$ to derive the scalar representation $W_{i}^{n} \in R^{m}$ from $\vec U_{i}^{1,n}$ and $\vec U_{i}^{2,n} \in R^{m \times 3}$. Within each GEDT layer, all neighboring atoms exchange information through the differential transformer. Scalar and vector representations also interact with each other. This scalar representation is then utilized to interact with the intermediate node scalar representation $y_{i}^{n}$:

\begin{equation}
w_{i}^{n}= \langle \vec U_{i}^{1,n},\vec U_{i}^{2,n}\rangle. 
\end{equation} 

Prior to the final output, we further convert $y_{i}^{n}$ into three different pieces:

\begin{equation}
O_{i}^{1,n}=W^{O^{1},n}y_{i}^{n}+b^{O^{1},n},
\end{equation}
\begin{equation}
O_{i}^{2,n}=W^{O^{2},n}y_{i}^{n}+b^{O^{2},n},
\end{equation}
\begin{equation}
O_{i}^{3,n}=W^{O^{3},n}y_{i}^{n}+b^{O^{3},n}.
\end{equation}

In the end, we present the updated node scalar representation $x_{i}^{n+1}$ and the node vector representation $\Vec{V}_{i}^{n+1}$ for the $(n+1)^{th}$ layer:

\begin{equation}
    x_{i}^{n+1}=x_{i}^{n}+\langle \vec U_{i}^{1,n},\vec U_{i}^{2,n}\rangle \odot O_{i}^{2,n}+O_{i}^{3,n},
\end{equation}
\begin{equation}
    \vec V_{i}^{n+1} = \vec V_{i}^{n}+O_{i}^{1,n} *\vec U_{i}^{3,n}+\vec u_{i}^{n}.
\end{equation}

As a result, both the node scalar and vector representations in the $(n+1)^{th}$ layer incorporate scalar and vector information from both nodes and edges in the $n^{th}$ layer. Once we reach the final GEDT layer (the n$^{th}$ layer below, and n=3 in this work), we compute NN-based per atom energy $E_{i}^{\text{NN}}$ based on $x_{i}^{n}$ by an MLP:

\begin{equation}
    E_{i}^{\text{NN}} = \text{MLP}(x_{i}^{n}),
\label{energymlp}
\end{equation} 
and we compute NN-based forces by:

\begin{equation}
    \Vec{f}_{ij}^{\text{ NN}} = -\frac{\partial E^{\text{NN}}}{\partial r_{ij}}+\frac{\partial E^{\text{NN}}}{\partial r_{ji}},
    \Vec{f}_{i}^{\text{ NN}}=\sum_{j\neq i}\Vec{f}_{ji}^{\text{ NN}}.
    \label{nnfij}
\end{equation}

Given the atomic structure, we can directly calculate the dispersion interactions by D3 dispersion correlation~\cite{2015_Schröder}. The final energy and forces predicted in this work are sum of the NN predicted quantities and the dispersion interactions.

\section{Initial Training}\label{subsec:si_dft}

In this work, we use DFT calculated energy, forces, and virial tensor of atomic clusters to train the machine learning force field. Here, atomic clusters were extracted from MD trajectories of organic liquids. The molecules included in the dataset are listed in the \textit{DFT + Density Alignment} and \textit{DFT only} sets in Figure \ref{fig:benchmark}. 

At the beginning, we use the DFT dataset in Ref. \cite{gong2024bamboo} with relevant molecules to train an ensemble of five GEDT models. This ensemble of GNNs is applied to predict the uncertainty of atoms in MD trajectories by the standard deviation of the prediction of forces from the ensemble. Then, we conducted MD simulation with NPT ensemble for 500 ps at 300 K with a randomly picked model on each pure organic solvent in the \ref{tab:si_data_density} with $true$ in column \textit{DFT included} to generate the trajectories for cluster extraction. Atoms in the trajectories with high predicted force uncertainty ($\sigma_{||\Vec{f}||} > 3$ kcal/(mol$\cdot$\AA)) by the ensemble GNNs are selected as the center of the specific clusters, where connected atoms and neighboring molecules are included during the identification of a cluster. More details of the construction of the DFT training set are provided in Ref. \cite{gong2024bamboo}.

Following the extraction, labels including energy, forces, and virials of atomic clusters were calculated with DFT using the open-source GPU4PySCF \cite{gpu4pyscf} software developed by ByteDance Inc. The calculation is performed with B3LYP~\cite{b3lyp} functional alongside the def2-svpd~\cite{svpd} basis set. The self-consistent field calculation is performed with a convergence criterion of $1.0\times10^{-10}$ a. u., allowing a maximum of 100 iterative cycles. 

After the first round of data collection, we use all the collected DFT data to train an ensemble of GEDT models. We repeat the force-uncertainty based active learning loop until all organic liquids can be stably simulated without spurious reactions during the NPT MD simulations. All organic liquids can remain stable during the MD simulations after 2 rounds of data collection. 

In this work, we train GEDT models with DFT labels of energy, forces, and virial tensors. The total loss is calculated as:

\begin{equation}
    \begin{split}
    L_{\text{training}} = \frac{1}{N_{\text{cluster}}} \sum_{k}^{N_{\text{cluster}}}\sum_{i}^{N_{\text{atom}}}&(\alpha_{\text{energy}}^{\text{training}} (E_{k}^{\text{DFT}} - E_{k}^{\text{BAMBOO-TC}})^{2} \\
         &+ \frac{1}{N_{\text{atom}}}\alpha_{\text{force}}^{\text{training}} ||\Vec{f}_{k,i}^{ \text{DFT}} - \Vec{f}_{k,i}^{ \text{BAMBOO-TC}}||^{2} \\
         &+ \alpha_{\text{virial}}^{\text{training}} ||\textbf{T}_{k}^{\text{DFT}} - \textbf{T}_{k}^{\text{BAMBOO-TC}}||^{2},
    \end{split}
    \label{trainingloss}
\end{equation}

where $\alpha_{\text{energy}}$ represents the weight for energy in the overall loss function, and so on. Here, we use the weights of 0.01, 0.3, and 0.01 for energy, forces, and virial, respectively. 

Specifically, the total number of training samples is 955,155, among which 859,634 are used for training and 95,521 for validation.

\section{Density Alignment}\label{subsec:si_alignment}

The details of the two-round density alignment implementation remain consistent with BAMBOO \cite{gong2024bamboo}, while the experimental data for the density alignment are different. The alignment process encompasses 8 molecules shown in the \textit{DFT + Density Alignment} set in Figure \ref{fig:benchmark}a. The experimental density data for these molecules are listed in Table~\ref{tab:si_data_density}.

The effect of density alignment is highlighted in Figure \ref{fig:benchmark}b, \ref{fig:benchmark}c and Table \ref{tab:thermal}. Comparing the predictive performance on each molecules, error on both density and thermal conductivity are reduced. The alignment process significantly decreases the density error for both the systems that were aligned and those that were not directly aligned. This enhancement indicates that density alignment, which modifies the strength of intermolecular interactions, can be somewhat transferable to molecules with similar structures that were not specifically included in the alignment. Also, this adjustment on the strength of intermolecular interactions also improves prediction results of thermal conductivity, indicating the successful reduction of systematic error between the DFT-calculated labels and the real world.

\section{Details of Molecular Dynamics}\label{sec:si_simulation}

We employ GROMACS~\cite{van2005gromacs} to generate the initial configuration of the molecules within the simulation box, and LAMMPS~\cite{thompson2022lammps} to conduct the molecular dynamics simulations for both GEDT and OPLS-AA. Each system begins with an energy minimization step, limited to a maximum of 1000 iterations and 100,000 evaluations. Following this, the liquid systems undergo equilibration in the NPT ensemble for 2 ns at 298K, using a time step of 1 fs. The density of the system is calculated as the average of the last 1ns of the NPT simulation. After equilibration, we perform a 1 ns production run in the NVT ensemble at 298K for further equilibration, using a time step of 1 fs. Finally, we create a temperature gradient along the z-axis and conduct a 2 ns production MD simulation in the NVE ensemble, using a time step of 0.5 fs, to evaluate the thermal conductivity of the system as in Ref.~\cite{muller1997rnemd}.

The thermal conductivity presented in this paper is calculated as follows. First, we divide the simulation box into 20 slabs along z-axis and calculate the mean kinetic temperature (MKT) of each slabs, as illustrated in Figure~\ref{fig:si_thermal}a. And kinetic energy will be swapped every 500 timesteps between the hottest atom in a selected cold slab and the coldest atom in a selected hot slab. This creates a well-defined steady-state heat flux. The temperature gradient is fit from the temperature profile of each slabs of the box, as illustrated in Figure~\ref{fig:si_thermal}b. Subsequently, the thermal conductivity can be calculated as:

\begin{equation}
    \kappa = \lim_{\partial{T}/\partial{z}\to0} \lim_{t\to\infty} - \frac{\langle J_z(t) \rangle}{\langle \partial{T}/\partial{z} \rangle}
    = - \frac{\sum \frac{m}{2} (v_h^2 - v_c^2)}{2tL_xL_y \langle \partial{T}/\partial{z} \rangle}
\end{equation}

where, $\langle J_z(t) \rangle$ represents the heat flux along the z-axis, the subscripts $h$ and $c$ denote the hot and cold particles, $m$ denotes the identical mass whose velocities are exchanged, $L_x$ and $L_y$ represent the box length on x and y axis, respectively. The temperature gradient of the system is presented as the mean of the two sections of the system. The $\kappa$ values of the systems are calculated using the gradient of the mean temperatures and the mean value of heat flux, of the last 1 ns of the NVE simulation.

\section{Organic Liquid Data}\label{subsec:si_data}

In this section, we list all the simulated and experimental properties related to this work for readers' reference in Table~\ref{tab:si_data_density} and Table ~\ref{tab:si_data_thermal}. 

\begin{table}
    \caption{Experimental and predicted data of density at 298K. The reported results are the mean values and the standard deviations over three independent simulations are used as error bars.}
    \label{tab:si_data_density}
    \centering
    \resizebox{\linewidth}{!}{
    \begin{tabular}{cc|cccccc}
        \hline
        Molecule & Chemical Name  & \makecell{Exp. \\$\rho (g/cm^3)$} &  \makecell{OPLS-AA \\$\rho (g/cm^3)$}&\makecell{BAMBOO-TC before \\alignment $\rho (g/cm^3)$}&\makecell{BAMBOO-TC after \\aligment $\rho (g/cm^3)$}& DFT& Alignment\\
        \hline
        MO&  methanol& 0.791~\cite{chen2002density_mo}& \makecell{0.649 (0.000483) }& \makecell{0.822 (0.000879) }& \makecell{0.791 (0.000293) }& \checkmark&\checkmark\\
        EO&  ethanol& 0.785~\cite{ortega1982density_eo}& \makecell{0.723 (0.000386)}& \makecell{0.801 (0.000255) }& \makecell{0.794 (0.000187) }& \checkmark&\checkmark\\
        TBUO&  tert-butanol& 0.775~\cite{erastova2017density_tbuo}& \makecell{0.776 (0.000228) }& \makecell{0.833 (0.000522) }& \makecell{0.776 (0.000582) }& \checkmark&\checkmark\\
 EF& \makecell{ethyl formate} & 0.914~\cite{ohta1980density_ef}& \makecell{0.879 (0.000354) }& \makecell{0.880 (0.000687) }& \makecell{0.913 (0.000478) }& \checkmark&\checkmark\\
 EA& \makecell{ethyl acetate}& 0.902~\cite{gong2024bamboo}& \makecell{0.819 (0.000193) }& \makecell{0.905 (0.000449) }& \makecell{0.902 (0.000515) }& \checkmark&\checkmark\\
 EIB& \makecell{ethyl iso-butyrate} & 0.865~\cite{byrne2018density_eib}& \makecell{0.810 (0.000176) }& \makecell{0.897 (0.000255) }& \makecell{0.868 (0.000322) }& \checkmark&\checkmark\\
 DMC& \makecell{dimethyl carbonate} & 1.069~\cite{gong2024bamboo}& \makecell{0.970 (0.000159) }& \makecell{1.028 (0.000943) }& \makecell{1.069 (0.000656) }& \checkmark&\checkmark\\
        \hline
        EP&  \makecell{ethyl propionate} & 0.884~\cite{sastry2013density_ep}& \makecell{0.795 (0.0000740) }& \makecell{0.904 (0.000878) }& \makecell{0.891 (0.000336) }& \checkmark&\\
        EB&  \makecell{ethyl butyrate}& 0.874~\cite{gong2024bamboo}& \makecell{0.792 (0.000267) }& \makecell{0.889 (0.00112) }& \makecell{0.892 (0.000414) }& \checkmark&\\
        IPRO&  \makecell{isopropyl-alcoho}& 0.781~\cite{gong2024bamboo}& \makecell{0.750 (0.000353) }& \makecell{0.812 (0.00156) }& \makecell{0.792 (0.000241) }& \checkmark&\\
        PC&  \makecell{propylene carbonate}& 1.204~\cite{gong2024bamboo}& \makecell{1.082 (0.000307) }& \makecell{1.244 (0.000937) }& \makecell{1.240 (0.000656) }& \checkmark&\\
 DOX& \makecell{dioxane}& 1.028~\cite{gong2024bamboo}& \makecell{0.798 (0.0114) }& \makecell{0.922 (0.000582) }& \makecell{0.982 (0.000421) }& \checkmark&\\
        \hline
        MA&  \makecell{methyl acetate} & 0.934~\cite{fao2024_ma_93}& \makecell{0.845 (0.000172) }& \makecell{0.934 (0.000278) }& \makecell{0.937 (0.000349) }& &\\
 MF& \makecell{methyl formate}& 0.967~\cite{gong2024bamboo}& \makecell{0.923 (0.000194) }& \makecell{0.895 (0.000777) }& \makecell{0.965 (0.00187) }& &\\
        PRA&  \makecell{propyl acetate} & 0.888~\cite{gardas2007density_pra_882}& \makecell{0.822 (0.0000480) }& \makecell{0.926 (0.000810) }& \makecell{0.886 (0.000379) }& &\\
        IPA&  \makecell{iso-pentyl acetate} & 0.876~\cite{wang2020density_ipa}& \makecell{0.717 (0.000301) }& \makecell{0.886 (0.000720) }& \makecell{0.903 (0.000416) }& &\\
        DEE&  \makecell{diethyl ether}& 0.713~\cite{crchandbook}& \makecell{0.537 (0.00255) }& \makecell{0.746 (0.000403) }& \makecell{0.722 (0.000872) }& &\\
 PRO& \makecell{1-propanol}& 0.803~\cite{crchandbook}& \makecell{0.739 (0.000847) }& \makecell{0.831 (0.000743) }& \makecell{0.779 (0.000567) }& &\\
        EV&  \makecell{ethyl valerate}& 0.874~\cite{density1}& \makecell{0.802 (0.000386) }& \makecell{0.891 (0.000722) }& \makecell{0.893 (0.000569) }& &\\
        EH&  \makecell{ethyl hexanoate}& 0.870~\cite{density1}& \makecell{0.820 (0.000405) }& \makecell{0.899 (0.000254) }& \makecell{0.881 (0.000861) }& &\\
        \hline
    \end{tabular}
    }
\end{table}

\begin{table}
    \caption{Experimental and predicted data of thermal conductivity at 298K. The reported results are the mean values and the standard deviations over three independent simulations are used as error bars.}
    \label{tab:si_data_thermal}
    \centering
    \resizebox{\linewidth}{!}{
    \begin{tabular}{cc|cccccc}
        \hline
         Molecule & Chemical Name  & \makecell{Exp. \\$\kappa  (W/mK)$} &  \makecell{OPLS-AA \\$\kappa (W/mK)$}&\makecell{BAMBOO-TC before\\alignment $\kappa (W/mK)$}&\makecell{BAMBOO-TC after \\alignment $\kappa (W/mK)$}& DFT& Alignment\\
        \hline
        MO& methanol& 0.200~\cite{tc_mo}& \makecell{0.307 (0.0148) }&  \makecell{0.244 (0.00439) }&\makecell{0.236 (0.0142) }& \checkmark&\checkmark\\
        EO& ethanol& 0.163~\cite{tc_eo}& \makecell{0.288 (0.0292) }&  \makecell{0.206 (0.00592) }&\makecell{0.212 (0.0134) }& \checkmark&\checkmark\\
        TBUO& tert-butanol& 0.116~\cite{yaws1997handbook1}& \makecell{0.230 (0.00321) }&  \makecell{0.195 (0.00692) }&\makecell{0.146 (0.00476) }& \checkmark&\checkmark\\
 EF& \makecell{ethyl formate} & 0.161~\cite{tc_ef_293}& \makecell{0.329 (0.00686) }& \makecell{0.136 (0.00264) }& \makecell{0.151 (0.00321) }& \checkmark&\checkmark\\
 EA& \makecell{ethyl acetate}& 0.143~\cite{yaws1995handbook2}& \makecell{0.265 (0.00107)}& \makecell{0.155 (0.00943) }& \makecell{0.140 (0.00438) }& \checkmark&\checkmark\\
 EIB& \makecell{ethyl iso-butyrate} & 0.127~\cite{yaws2009tc_eib}& \makecell{0.243 (0.00336)}& \makecell{0.159 (0.00592) }& \makecell{0.133 (0.00309) }& \checkmark&\checkmark\\
 DMC& \makecell{dimethyl carbonate} & 0.163~\cite{yaws1997handbook1}& \makecell{0.312 (0.00545) }& \makecell{0.141 (0.00300) }& \makecell{0.162 (0.00137) }& \checkmark&\checkmark\\
        \hline
        EP& \makecell{ethyl propionate} & 0.133~\cite{yaws1995handbook2}& \makecell{0.271 (0.00328) }&  \makecell{0.161 (0.00347) }&\makecell{0.154 (0.00393) }& \checkmark&\\
 EB& \makecell{ethyl butyrate}& 0.137~\cite{yaws1997handbook1}& \makecell{0.251 (0.0125) }& \makecell{0.171 (0.00290) }& \makecell{0.160 (0.00581) }& \checkmark&\\
 IPRO& \makecell{isopropyl-alcoho}& 0.135~\cite{yaws2009tc_eib}& \makecell{0.243 (0.00259) }& \makecell{0.192 (0.00574) }& \makecell{0.171 (0.00732) }& \checkmark&\\
 PC& \makecell{propylene carbonate}& 0.165~\cite{yaws1995handbook2}& \makecell{0.250 (0.00111) }& \makecell{0.200 (0.000572) }& \makecell{0.200 (0.00134) }& \checkmark&\\
 DOX& \makecell{dioxane}& 0.162~\cite{yaws1997handbook1}& \makecell{0.201 (0.00496) }& \makecell{0.131 (0.00229) }& \makecell{0.154 (0.00219) }& \checkmark&\\
        \hline
        MA& \makecell{methyl acetate} & 0.155~\cite{yaws1997handbook1}& \makecell{0.295 (0.00351) }&  \makecell{0.141 (0.00185) }&\makecell{0.140 (0.000163) }& &\\
        MF& \makecell{methyl formate}& 0.187~\cite{yaws1995handbook2}& \makecell{0.378 (0.0193) }&  \makecell{0.124 (0.000825) }&\makecell{0.152 (0.00179) }& &\\
        PRA& \makecell{propyl acetate} & 0.140~\cite{yaws1995handbook2}& \makecell{0.256 (0.00765) }&  \makecell{0.160 (0.00186) }&\makecell{0.140 (0.00456) }& &\\
        IPA& \makecell{iso-pentyl acetate} & 0.140~\cite{yaws1997handbook1}& \makecell{0.202 (0.00906) }& \makecell{0.149 (0.00329) }& \makecell{0.158 (0.00238) }& &\\
        DEE&  \makecell{diethyl ether}& 0.130~\cite{yaws1997handbook1}& \makecell{0.190 (0.00697) }& \makecell{0.179 (0.00300) }& \makecell{0.146 (0.00376) }& &\\
 PRO& \makecell{1-propanol}& 0.168~\cite{yaws1997handbook1}& \makecell{0.249 (0.0109) }& \makecell{0.229 (0.00969) }& \makecell{0.190 (0.00410) }& &\\
        EV&  \makecell{ethyl valerate}& 0.136~\cite{thermalcond}& \makecell{0.274 (0.00809) }& \makecell{0.175 (0.00183) }& \makecell{0.168 (0.000586) }& &\\
        EH&  \makecell{ethyl hexanoate}& 0.138~\cite{thermalcond}& \makecell{0.268 (0.0126) }& \makecell{0.174 (0.0112) }& \makecell{0.169 (0.00332) }& &\\
        \hline
    \end{tabular}
    }
\end{table}

\end{document}